\documentclass[a4paper,USenglish,cleveref,thm-restate,numberwithinsect]{lipics-v2021}

\hideLIPIcs
\nolinenumbers

\author{Léo Stefanesco}{MPI-SWS}{}{}{}
\author{Azalea Raad}{Imperial College}{}{}{}
\author{Viktor Vafeiadis}{MPI-SWS}{}{}{}

\authorrunning{Stefanesco, Raad and Vafeiadis}


\usepackage{defs2}

\usepackage[notion,
  paper,
hyperref,quotation,xcolor]{knowledge}

\knowledgeconfigure{protect quotation={tikzcd}}

\colorlet{linkcolor}{nicepurple}

\IfKnowledgePaperModeTF
{
  \knowledgestyle{intro notion}{italic}
  \knowledgestyle{notion}{color=black}
}{}
\IfKnowledgeCompositionModeTF
{
  \knowledgestyle{intro notion}{boldface,color=linkcolor}
  \knowledgestyle{notion}{color=linkcolor}
}{}
\IfKnowledgeElectronicModeTF
{
  \knowledgestyle{intro notion}{boldface,color=linkcolor}
  \knowledgestyle{notion}{color=linkcolor}
}{}

\knowledge{universal projection}{notion}
\knowledge{existential projection}{notion}
\knowledge{canonical projection}{notion}
\knowledge{method invocation}[method invocations]{notion}
\knowledge{method call}[method calls]{notion}
\knowledge{library interface}[library interfaces]{notion}
\knowledge{library}[libraries]{notion}
\knowledge{well-formed}[well-formedness]{notion}
\knowledge{well-formed implementation}{notion}
\knowledge{immediately well-formed}[immediate well-formedness]{notion}
\knowledge{consistent}[consistency]{notion}
\knowledge{encapsulated}[consistency]{notion}
\knowledge{total execution}[total executions]{notion}
\knowledge{total plain execution}{notion}
\knowledge{locally correct}[local correctness]{notion}
\knowledge{correct}[correctness]{notion}
\knowledge{hereditarily correct}[hereditary correctness]{notion}
\knowledge{hereditarily well-formed}[hereditary well-formedness]{notion}
\knowledge{valid execution}[valid executions]{notion}
\knowledge{locally sound}{notion}
\knowledge{execution chain}[Execution chains|execution chains]{notion}
\knowledge{plain execution}[plain executions]{notion}
\knowledge{plain execution chain}[plain execution chains]{notion}
\knowledge{local modular soundness}{notion}
\knowledge{global modular soundness}{notion}
\knowledge{execution}[executions]{notion}
\knowledge{pomset}[pomsets]{notion}
\knowledge{sequential specification}{notion}
\knowledge{linearizable}{notion}
\knowledge{durably linearizable}{notion}
\knowledge{implements}{notion}
\knowledge{maximal prefix}{notion}
\knowledge{prefix}{notion}
\knowledge{complete operation}[complete operations]{notion}
\knowledge{incomplete operation}[incomplete operations]{notion}
\knowledge{era}[eras]{notion}
\knowledge{sequential composition}{notion}
\knowledge{down-closed implementation}{notion}
\knowledge{down-closed library}{notion}
\knowledge{implementation}[implementations]{notion}

\graphicspath{{../}}

\title{Specifying and Verifying Persistent Libraries}

\usepackage[square,numbers]{natbib}
\begin{document}
\maketitle

\begin{abstract}
  We present a general framework for specifying and verifying \emph{persistent
  libraries}, that is, libraries of data structures that provide some persistency
guarantees upon a failure of the machine they are executing on.
Our framework enables modular reasoning about the correctness of individual
libraries (horizontal and vertical compositionality)
and is general enough to encompass all existing persistent library specifications
ranging from hardware architectural specifications
to correctness conditions such as durable linearizability.
As case studies, we specify the \Flit and \Mirror libraries, verify their
implementations over \pxes, and use them to build higher-level durably
linearizable libraries, all within our framework.
We also specify and verify a persistent transaction library that highlights
some of the technical challenges which are specific to persistent memory compared
to weak memory and how they are handled by our framework.

\end{abstract}


\section{Introduction}

Persistent memory (PM), also known as non-volatile memory (NVM), is a new kind
of memory, which can be used to extend the capacity of regular RAM, with the
added benefit that its contents are preserved after a crash (\eg a power failure).
Employing PM can boost the performance of any program with access to data that
needs to survive power failures, be it a complex database or a plain text
editor.

Nevertheless, doing so is far from trivial.
Data stored in PM is mediated through the processors' caching hierarchy, which
generally does not propagate all memory accesses to the PM in the order issued
by the processor, but rather performs these accesses on the cache and only
propagates them to the memory asynchronously when necessary (\ie upon a cache
miss or when the cache has reached its capacity limit).
Caches, moreover, do not preserve their contents upon a power failure, which
results in rather complex persistency models describing when and how stores
issued by a program are guaranteed to survive a power failure.
To ensure correctness of their implementations, programmers have to use
low-level primitives, such as \emph{flushes} of individual cache lines,
\emph{fences} that enforce ordering of instructions, and \emph{non-temporal
  stores} that bypass the cache hierarchy.

These primitives are often used to implement higher-level abstractions,
packaged into \emph{persistent libraries}, \ie collections of data structures
that must guarantee to preserve their contents after a power failure.
Persistent libraries can be thought of as the analogue of concurrent libraries
for persistency.
And just as concurrent libraries require a specification, so do persistent
libraries.

The question naturally arises: what is the right specification for persistent
libraries? Prior work has suggested a number of candidate definitions, such as
\emph{durable linearizability}, \emph{buffered durable linearizability}
\cite{Izraelevitz}, and \emph{strict linearizability} \cite{strict-lin}, which
are all extensions of the well-known correctness condition for concurrent data
structures (\ie linearizability \cite{linearisability}). In general, these
definitions stipulate the existence of a total order among all executed library
operations, a contiguous prefix of which is persisted upon a crash: the various
definitions differ in exactly what this prefix should be, \eg whether it is
further constrained to include all fully executed operations.

Even though these specifications have a nice compositionality property, we argue
that none of them are \emph{the} right specification pattern for \emph{every}
persistent concurrent library.
While for high-level persistent data structures, such as stacks and queues, a
strong specification such as durable or strict linearizability would be most
appropriate, this is certainly not the case for a collection of low-level
primitives.
Take, for instance, a library whose interface simply exposes the exact
primitives of the underlying platform: memory accesses, fences and flushes.
Their semantics, recently formalized in \cite{Px86, TamingPx86, NTSx86} in the
case of the \intelname architecture and in \cite{parm-oopsla,parmv8} in the
case of the \armname architecture, quite clearly do not fit into the framework
of the durable linearizability definitions.
More generally, there are useful concurrent libraries (especially in the
context of weak memory consistency) that are not linearizable \cite{yacovet};
it is, therefore, conceivable that making those libraries persistent will
require weak specifications.

Another key problem with attempting to specify persistent libraries
\emph{modularly} is that they often break the usual abstraction boundaries.
Indeed, some models such as epoch persistency \cite{persist-buffering,pelley-persistency}
provide a global persistency barrier that affects all memory locations, and
therefore all libraries using them.
Such global operations also occur at higher abstraction layers: persistent
transactional libraries often require memory locations to be registered with the library
in order for them to be used inside transactions.  As such, to ensure compatibility
with such transactional libraries, implementers of other libraries must
register all locations they use and ensure that any component libraries they use
do the same.

In this paper, we introduce a \emph{general declarative framework} that
addresses both of these challenges.
Our framework provides a very flexible way of specifying persistent
libraries, allowing each library to have a very different specification---be it
durable linearizability or a more complex specification in the style of the
hardware architecture persistency models.
Further, to handle libraries that have a global effect (such as persistent
barriers above) or, more generally, that make some assumptions about the
internals of all other libraries, we introduce a \emph{tag} system, allowing us
to describe these assumptions \emph{modularly}.

Our framework supports both \emph{horizontal} and \emph{vertical compositionality}. That is, we can verify an execution containing multiple
libraries by verifying each library separately (horizontal
compositionality). Moreover, we can completely verify 
the implementation of a library over a set of other libraries using the
specifications of its constituent libraries without referring to their
implementations (vertical compositionality). To achieve the latter, we define a
semantic notion of substitution in terms of execution graphs, which replaces
each library node by a suitably constrained set of nodes (its implementation).

For simplicity, in \cref{sec:overview}, we develop a first version of our
framework over the classical notion of an execution \emph{history}
\cite{linearisability}, which we extend with a notion of crashes.
This basic version of our framework includes full support for weak persistency
models but assumes an interleaving semantics of concurrency; \ie sequential
consistency (SC) \cite{lamport-sc}.

Subsequently, in \cref{sec:framework} we generalise and extend our framework to handle weak consistency models such as x86-TSO \cite{tso}  and RC11 \cite{rc11},
thereby allowing us to represent hardware persistency models such as \pxes \cite{Px86}
and \parm \cite{parm-oopsla}, in our framework.
To do so, we rebase our formal development over execution graphs
using \yacovet \cite{yacovet} as a means of specifying the consistency
properties of concurrent libraries.

We illustrate the utility of our framework by encoding in it a number of
existing persistency models, ranging from actual hardware models such as \pxes
\cite{Px86}, 
to general-purpose correctness conditions
such as durable linearizability \cite{Izraelevitz}.
We further consider two case studies, chosen to demonstrate the expressiveness
of our framework beyond the kind of case studies that have been worked out
in the consistency setting.

%
First, in \cref{sec:flit_and_mirror} we use our framework to develop the first formal specifications of the \Flit \cite{Flit} and \Mirror \cite{Mirror} libraries and establish the correctness of not only their implementations against their respective specifications, but also their associated constructions for turning a linearizable library into a durably linearizable
one. This generic theorem is new compared to the case studies in
\cite{yacovet}, and leverages our `semantic' approach in \cref{sec:framework}.
Moreover, our proofs of these constructions are the \emph{first} to establish this result in a weak consistency setting. 

Second, in \cref{sec:transactions} we specify and prove an implementation of a
persistent transactional library~$\Ltrans$, which provides a high-level construction to persist a set of writes \emph{atomically}.
The $\Ltrans$ library illustrates the need for a `well-formedness' specification (in
addition to its consistency and persistency specifications) that requires clients
of the $\Ltrans$ library to ensure \eg that $\Ltrans$ writes appear only inside transactions.
Moreover, it demonstrates the use of our tagging system to enable other libraries
to interoperate with it.

\smallskip\noindent \paragraph{Contributions and Outline}
The remainder of this article is organised as follows.
\begin{enumerate}
	\item[\textbf{\cref{sec:overview}}]  We present our general framework for specifying and verifying persistent libraries in the strong sequential consistency setting. 
	\item[\textbf{\cref{sec:framework}}] We further generalise our framework to account for weaker consistency models.
	\item[\textbf{\cref{sec:flit_and_mirror}}] We use our framework to develop the first formal specifications of the \Flit and \Mirror libraries, verify their implementations against their specifications and prove their general construction theorems for turning linearizable libraries to durably linearizable ones. 
	\item[\textbf{\cref{sec:transactions}}] We specify a persistent transactional library $\Ltrans$, develop an implementation of $\Ltrans$ (over the Intel-x86 architecture) and verify it against its specification. We then consider two case studies of vertical and horizontal composition in our framework using $\Ltrans$.
\end{enumerate}
We conclude and discuss related and future work in \cref{sec:conclusions}.
The full proofs of all theorems stated in the paper are given in the technical appendix.

\section{A General Framework for Persistency}
\label{sec:overview}
We present our framework for specifying and verifying persistent \emph{libraries},
which are collections of methods that operate on durable data structures. 
Following Herlihy \etal \cite{linearisability}, we will represent program
histories over a collection of libraries $\coll$ as $\coll$-histories,
\ie as sequences of calls to the methods of $\coll$,
which we will then gradually enhance to model persistency semantics.
Throughout this section, we assume an underlying sequential consistency semantics;
in \cref{sec:framework} we will generalize our framework to account for weaker consistency models. 

In the following, we assume the following infinite domains:
$\MName$ of method names, $\Loc$ of memory locations,
$\ThreadIds$ of thread identifiers, and $\Val \supseteq \Loc\cup\ThreadIds$ of values.
We let $m$ range over method names, $x$ over memory locations, $\thread$ over thread identifiers, and $v$ over values.
An optional value~$\valbot \in \Valbot$ is either a value~$v \in \Val$ or~$\bot
\notin \Val$.

\subsection{Library Interfaces}
\label{sec:overview-libinterface}

A \emph{library interface} declares a set of method invocations of the form $m(\vec v)$.
Some methods are are designated as constructors; 
a constructor returns a location pointing to the new library instance (object),
which is passed as an argument to other library methods.
An interface additionally contains a function, $\loc$, which extracts these
locations from the arguments and return values of its method calls.
\begin{definition}\label{def:library-interface-v1}
  A \emph{library interface} $\genlibint$ is a tuple \(\tuple{\methods, \methodsconstr,
  \loc}\), where $\methods \subseteq \pset{\MName \times \Val^*}$ is the set of method invocations,
  $\methodsconstr \subseteq \methods$ is the set of constructors, and $\loc :
  \methods \times \Val_\bot \rightarrow \pset{\Loc}$ is the location function.
\end{definition}

\begin{example}[Queue library interface]
\label{example:queue-interface}
  The queue library interface, $\Lqueue$, has three methods:
  a constructor $\QueueNew()$, which returns a new empty queue;
  $\QueuePush(x, v)$ which adds value~$v$ to the end of queue~$x$; and
  $\QueuePop(x)$ which removes the head entry in queue~$x$. 
  We define $\loc(\QueueNew(), x) =  \loc(\QueuePush(x, \_), \_) = \loc(\QueuePop(x), \_) = \{ x \}$.
\end{example}

A \emph{collection} $\coll$ is a set of library interfaces with disjoint method names.
When $\coll$ consists of a single library interface $\genlibint$, we often write $\genlibint$ instead
of $\{\genlibint\}$.

\subsection{Histories}
\label{sec:overview-histories}

Given a collection $\coll$, an event $e \in \Events(\coll)$ of $\coll$ is either
a method invocation $m(\vec v)_\thread$ with $m(\vec v) \in \bigcup_{\genlibint \in \coll} \genlibint{.}\methods$ and~$\thread \in \ThreadIds$
or method response (return) event $\ret(v)_\thread$.

A $\coll$-history is a sequence of events of $\coll$ whose projection to each
thread is an alternating sequence of invocation and return events which starts
with an invocation.

\begin{definition}[Sequential event sequences]
A sequence of events $e_1 \ldots e_n$ is \emph{sequential}
if all its odd-numbered events $e_1,e_3,\ldots$ are invocation events
and all its even-numbered events $e_2,e_4,\ldots$ are return events.
\end{definition}

\begin{definition}[Histories]
  A $\coll$-\emph{history} is a finite sequence of events $\hist \in \Events(\coll)^*$,
  such that for every thread $\thread$,
  the sub-sequence $\hist[\thread]$ comprising only of $\thread$ events is sequential.
  The set $\Hist{\coll}$ denotes the set of all $\coll$-histories.
\end{definition}

\noindent
When clear from the context, we refer to \emph{occurrences} of events in a history by their corresponding events.
For example, if $\hist = e_1 \dots e_n$ and $i < j$,
we say that $e_i$ \emph{precedes}~$e_j$ and that $e_j$ \emph{succeeds} $e_i$.
Given an invocation~$m(\vec v)_\thread$ in~$\hist$, its \emph{matching return} (when it exists) is the first event of the form~$\ret(v)_\thread$ that succeeds it (they share the same thread).
A \emph{call} is a pair~$\genmethlabthread$ of an invocation and either its matching
return~$\valbot \in \Val$ (\emph{complete call}) or~$\valbot = \bot$ (\emph{incomplete call}).

A \emph{library} (specification) comprises an interface and a set of \emph{consistent} histories.
The latter captures the allowed behaviors of the library,
which is a guarantee made by the library implementation.
\begin{definition}\label{def:library-specification-v1}
  A \emph{library specification} (or simply a \emph{library})~$\genlib$ is a tuple
  $\tuple{\genlibint, \libConsistent}$,
  where
  $\genlibint$ is a library interface, and
  $\libConsistent \suq \Hist{\genlibint}$ denotes its set of \emph{consistent} histories.
\end{definition}

\subsection{Linearizability}
\label{sec:overview-lin}

Linearizability~\cite{linearisability} is a standard way of
specifying concurrent libraries that have a sequential specification~$S$,
denoting a set of finite sequences of complete calls.
Given a sequential specification~$S$, a concurrent library~$\genlib$ is
linearizable under $S$ if each consistent history of~$\genlib$ can be
\emph{linearized} into a sequential one in~$S$, while respecting the
\emph{happens before} order, which captures causality between calls.
It is sufficient to consider consistent executions because inconsistent
executions are, by definition, guaranteed by the library to never happen.
Happens-before is defined as follows.
\begin{definition}[Happens-Before]
  A method call $C_1$ \emph{happens before} another method call $C_2$ in a history $\hist$, written
  $C_1 \happensbefore_\hist C_2$ if the response of $C_1$ precedes the invocation of $C_2$ in $\hist$.
  When the choice of $\hist$ is clear from the context, we drop the $\hist$ subscript from $\happensbefore$.
\end{definition}

A history $\hist$ is \emph{linearizable} under a sequential specification~$S$
if there exists a linearization (in the order-theoretic sense) of
$\happensbefore_\hist$ that belongs to~$S$. The subtlety is the treatment
of incomplete calls, which may or may not have taken effect.
We write~$\compl(\hist)$ for the set of histories obtained from a history $\hist$ by appending zero or more matching return events. 
We write $\trunc(\hist)$ for the history obtained from $\hist$ by removing its incomplete calls.
We can then define linearizability as follows~\citep{HerlihyShavit}.
\begin{definition}
  A sequential history $\hist_\ell$ is a \emph{sequentialization} of a history $\hist$
  if there exists $\hist' \in \trunc(\compl(\hist))$ such
  that $\hist_\ell$ is a linearization of $\happensbefore_{\hist'}$.
  A history $\hist$ is \emph{linearizable under} $S$ if there exists a
  sequentialization of $\hist$ that belongs to~$S$.
  A library $\genlib$ is \emph{linearizable} under $S$ if all its consistent
  histories are linearizable under $S$.
\end{definition}

For instance, we can specify the notion of \emph{linearizable queues} as those
that linearizable under the following sequential queue specification, $S_\Queue$.

\begin{example}[Sequential queue specification]
\label{example:sequential-queue}
  The behaviors of a sequential queue, $S_\Queue$, is expressed as a set of sequential
  histories as follows. 
  Given a history~$\hist$ of~$\genlibint_\Queue$ and a location~$x \in \Loc$, let $\hist[x]$ denote the sub-history containing calls~$c$ such that~$\loc(c) = \{x\}$.
  We define~$S_\Queue$ as the set of all sequential histories~$\hist$
  of~$\genlibint_\Queue$ such that for all~$x \in \Loc$, $\hist[x]$ is of the form
 $
    \methlab{\QueueNew}{}{x}{\thread_0}
    \;
    e_1 \; \cdots \; e_n
  $,  
  where each $\QueuePop$ call in $e_1 \; \cdots \; e_n$ returns the value
  of the $k$-th $\QueuePush$ call, if it exists and precedes the $\QueuePop$,
  where $k$ is the number of preceding $\QueuePop$ calls returning non-null values;
  otherwise, it returns null.
\end{example}

\subsection{Adding Failures}
\label{sec:overview-failures}

Our framework so far does not support reasoning about persistency as it lacks
the ability to describe the persistent state of a library after a failure. Our
first extension is thus to extend the set of events of a collection,
$\Events(\coll)$, with another type of event, a crash event $\Crash$.

Crash events allow us to specify the durability guarantees of a library. For
instance, a library that does not persist any of its data may specify that a
history with crash events is consistent if all of its sub-histories between
crashes are (independently) consistent. In other words, in such a library, the
method calls before a crash have no effect on the consistency of the history
after the crash.
We modify the definition of happens-before accordingly by treating it both as an
invocation and a return event. We also assume that, after a crash, the thread
ids of the new threads are distinct from that of all the pre-crash threads.
For libraries that do persist their data, a useful generic specification is
\emph{durable linearizability}~\citep{Izraelevitz}, defined as follows.

\begin{definition}
  Given a history~$\hist$, let $\ops(\hist)$ denote the sub-history obtained
  from $\hist$ by removing all its crash markers.
  A history $\hist$ is \emph{durably linearizable} under $S$ if there exists a sequentialization $\hist_\ell$ of $\ops(\hist)$ such that~$\hist_\ell \in S$.
\end{definition}
Intuitively, this ensures that operations persist before they return, and they
persist in the same order as they take effect before a crash.

Although durable linearizability can specify a large range of persistent data-structures, it can be too strong. 
For example, consider a (memory) register library~$\Lweakreg$ that only guarantees that writes to the \emph{same} location are persisted in the order they are observed by concurrent reads.
The $\Lweakreg$ methods comprise $\RegNew()$ to allocate a new register, $\RegWrite(x, v)$ to write $v$ to register $x$, and $\RegRead(x)$ to read from
register $x$.
The sequential specification~$\Sweakreg$ is simple: once a register is
allocated, a read $R$ on $x$ returns the latest value written to $x$, or $0$ if $R$ happens before all writes.
The associated durable linearizability specification requires that writes be
persisted in the linearization order; however, this is often not the case on
existing hardware, \eg in \pxes (the Intel-x86 persistency model)~\citep{Px86}.

A more relaxed and realistic specification would consider two linearizations of
the events: the standard \emph{volatile} order~$\klin$ and a \emph{persistent}
order~$\knvo$ expressing the order in which events are persisted.
The next sections will handle this more refined model, this paragraph only
gives a quick tastes of the kind of models that are implemented by hardware.
To capture the same-location guarantees, we stipulate a per-location ordering on
writes that is respected by both linearizations. Specifically, we require an
ordering~$\kmo$ of the write calls such that for all locations $x$:
\begin{enumerate*}[label=\arabic*)]
	\item restricting $\kmo$ to $x$, written  $\kmo_x$, totally orders writes to~$x$; and 
	\item $\kmo_x \subseteq \klin$ and~$\kmo_x \subseteq \knvo$.
\end{enumerate*}
Given a history~$\hist$, we can then combine these two linearizations by
using~$\klin$ after the last crash and $\knvo$ before.

Formally, a history $\hist$ with $n{-}1$ crashes can be decomposed into $n$ (crash-free) \emph{eras}; \ie $\hist = \hist_1 \cdot \Crash \cdots \Crash \cdot \hist_n$ where each $\hist_i$ is crash-free.
Let us write $\klin_i$ for $\klin \cap (\hist_i \times \hist_i)$ and so forth.
We then consider $k$-sequentializations of the form
\(
  \hist^k_\ell = \hist_\ell^{(1)} \cdots \hist_\ell^{(k-1)} \cdot \hist_\ell^{(k)}
\), 
where $\hist_\ell^{(k)}$ is a sequentialization of~$E_k$ \wrt $\klin_k$ and
$\hist^{(i)}_\ell$ is a sequentialization of~$E_i$ \wrt $\knvo_i$, for $i<k$.
We can now specify our weak register library as follows, where $\hist$ comprises $n$ eras:
\begin{center}
$
\hist \in
\Lweakreg{.}\libConsistent \iff \for{k \leq n}\exsts{\hist^k_\ell
\text{ $k$-seq. of $\hist$}} \hist^k_\ell \in \Sweakreg
$
\end{center}

\begin{example}\label{example:weak-register-history}
The following history is valid according to this specification
but not according to the durably linearizable one:
\begin{center}\footnotesize$
  W_{\thread_1}(x, 1) \cdot
  W_{\thread_2}(y,1) \cdot
  R_{\thread_3}(y) \cdot
  \ret_{\thread_3}(1)\cdot
  R_{\thread_3}(x)\cdot
  \ret_{\thread_3}(0) \cdot
  \Crash \cdot 
  R_{\thread_4}(y)\cdot
  \ret_{\thread_4}(0)\cdot
  R_{\thread_4}(x)\cdot
  \ret_{\thread_4}(1)
$\end{center}
While the writes to~$x$ ($W_{\thread_1}(x, 1)$) and~$y$ ($W_{\thread_2}(y, 1)$) are executing, thread~$\thread_3$ observes the new value (1) of~$y$ but the old value (0) of~$x$; \ie $\klin$ must order $W_{\thread_2}(y, 1)$ before $W_{\thread_1}(x, 1)$.
By contrast, after the crash the new value (1) of $x$ but the old value of $y$ (0) is visible; \ie $\knvo$ must order the two writes in the opposite order to~$\klin$ ($W_{\thread_1}(x, 1)$ before $W_{\thread_2}(y, 1)$).
\end{example}

\paragraph{Persist Instructions}
The persistent registers described above are too weak to be practical, as there
is no way to control how writes to different locations are persisted.
In realistic hardware models such as~\pxes, this control is afforded to the programmer using per-location \emph{persist} instructions (\eg \texttt{CLFLUSH}), ensuring that all writes on a location $x$ persist before a write-back on $x$.
Here, we consider a coarser (stronger) variant, denoted by \lstinline|PFENCE|, that ensures that \emph{all} writes (on \emph{all} locations) that happen before a \lstinline|PFENCE| are persisted.
Later in \cref{sec:framework} we describe how to specify the behavior of per-location persist operations.

Formally, we specify  \lstinline|PFENCE| by extending the specification of
$\Lweakreg$ with as follows: given history $\hist$, write call $c_w$ and \lstinline|PFENCE| event $c_f$, if $c_w \happensbefore_\hist c_f$, then $(c_w, c_f) \in \knvo$.
\begin{example}
  Consider the history obtained from \cref{example:weak-register-history} by adding a \lstinline|PFENCE|:
\begin{center}\footnotesize$
  W_{t_1}(x, 1) \cdot
  W_{t_2}(y,1) \cdot
  R_{t_3}(y) \cdot
  \ret_{t_3}(1)\cdot
  R_{t_3}(x)\cdot
  \ret_{t_3}(0) \cdot
  \text{\lstinline|PFENCE|}_{t_4}()\cdot
  \ret_{t_4}() \cdot
  \Crash \cdot 
  R_{t_4}(y)\cdot
  \ret_{t_4}(0)\cdot
  R_{t_4}(x)\cdot
  \ret_{t_4}(1)
$\end{center}
This history is no longer consistent according to the extended specification of
$\Lweakreg$: as \lstinline|PFENCE| has completed (returned), all its
$\happensbefore$-previous writes must have persisted and thus must be visible
after the crash (which is not the case for $W_{t_2}(y,1)$). 
\end{example}

\subsection{Adding Well-formedness Constraints}

Our next extension is to allow library specifications to constrain the
\emph{usage} of the library methods by the client of the library.
For example, a library for a mutual exclusion lock may require that the ``release lock'' method
is only called by a thread that previously acquired the lock and has not released it in between.
Another example is a transactional library, which may require that transactional read and write methods
are only called within transactions, \ie between a ``transaction-begin'' and a ``transaction-end'' method call.

We call such constraints library \emph{well-formedness} constraints,
and extend the library specifications with another component,
$\libWellformed\suq \Hist{\genlibint}$, which records the set of well-formed histories of the library.
Ensuring that a program produces only well-formed histories of a certain
library is an obligation of the clients of that library,
so that the library implementation can rely upon well-formedness being satisfied.

\subsection{Tags and Global Specifications}\label{sec:tags-and-global-specs}
The goal of our framework is not only to specify libraries in isolation, but also
to express how a library can enforce persistency guarantees across other libraries.
%
For example, consider a library~$\Ltrans$ for persistent transactions, where all operations wrapped within a transaction persist together \emph{atomically}; \ie either all or none of the operations in a transaction persist. 

The $\Ltrans$ methods are: 
\lstinline|PTNewReg| to allocate a register that can be accessed (read/written) within a transaction;
\lstinline|PTBegin| and \lstinline|PTEnd| to start and end a transaction, respectively; 
\lstinline|PTRead(x)| and \lstinline|PTWrite(x, v)| to read from and write to $\Ltrans$ register \lstinline|x|, respectively; and
\lstinline|PTRecover| to restore the atomicity of transactions whose histories were interrupted by a crash. 

Consider the snippet below, where the \lstinline|PEnq(q, 33)| (enqueuing \lstinline|33| into persistent queue~\lstinline|q|) and \lstinline|PSetAdd(s, 77)| (adding \lstinline|77| to persistent set \lstinline|s|) are wrapped within an $\Ltrans$ transaction and thus should take effect atomically and at the latest after
the end of the call to~\lstinline|PTEnd|.
\begin{lstlisting}[xleftmargin=1em]
PTBegin();
  PEnq(q, 33);
  PSetAdd(s, 77);
PTEnd();
\end{lstlisting}
Such guarantees are not offered by existing hardware primitives \eg on Intel-x86 or ARMv8 \citep{Px86,parm-oopsla} architectures.
%
As such, to ensure atomicity, the persistent queue and set implementations cannot directly use hardware reads/writes; rather, they must use those provided by the transactional library whose implementation could use \eg an undo-log to provide atomicity.

Our framework as described so far cannot express such cross-library persistency guarantees.
The difficulty is that the transactional library relies on other libraries using certain primitives. This, however, is against the spirit of \emph{compositional specification}, which precludes the transactional library from referring to other libraries (\eg the queue or set libraries).
Specifically, there are two challenges. 
First, both well-formedness requirements and consistency
guarantees of $\Ltrans$ must apply to \emph{any} method call that is designed
to use (transitively) the primitives of~$\Ltrans$.
Second, we must formally express atomicity (``all operations
persist atomically''), without $\Ltrans$ knowing what it means for a method of an arbitrary library to persist.
In other words, $\Ltrans$ needs to introduce an abstract notion of `having
persisted' for an operation, and guarantee that all methods in a transaction `persist' atomically.

To remedy this, we introduce the notion of \emph{tags}. Specifically, to address
the first challenge, the transactional library provides the tag~$\Ttag$ to
designate those operations that are `transaction-aware' and as such must be used
inside a transaction. To address the second challenge, the transaction library
provides the tag~$\perTag$, denoting an operation that has abstractly persisted.
The specification of~$\Ltrans$ then guarantees that all operations tagged
with~$\Ttag$ inside a transaction persist atomically, in that either they are
all tagged with~$\perTag$ of none of them are. Dually, using the well-formedness
condition, $\Ltrans$~requires that all operations tagged with~$\Ttag$ appear
inside a transaction.
Note that as the persistent queue and set libraries tag their operations with~$\Ttag$, verifying their implementations incurs related proof obligations; we will revisit this later when we formalize the notion of library implementations.

\begin{remark}[Why bespoke persistency?]
  The reader may question why `having persisted' is not a primitive
  notion in our framework, as in an existing model of \pxes~\citep{TamingPx86}
  where histories track the set~$P$ of persisted events.
  This is because associating a Boolean (`having persisted') flag with an operation may not be sufficient to describe whether it has persisted.
  To see this, consider a library $\Lpair$ with operations~$\PairWrite(x, l, r)$
  (writing $(l, r)$ to pair $x$), $\PairReadL(x)$ and $\PairReadR(x)$ (reading the left and right components of $x$, respectively).
  Suppose $\Lpair$ is implemented by storing the left component in an
  $\Ltrans$ register and the right component in a ~$\Lweakreg$ register.
  The specification of~$\Lpair$ would need to track the persistence of
  each component separately, and hence a single set~$P$ of persisted events
  would not suffice.
\end{remark}

Let us see how libraries can use these tags in \emph{global well-formedness and
consistency specifications}.
The dilemma is, on the one hand, the specification of~$\Ltrans$ needs to refer
to events from other libraries, but on the other hand, it should not depend on
other libraries to preserve encapsulation.
Our idea is to \emph{anonymize} these external events such that the global
specification depends only on their relevant tags.
A library should only rely on the tags it introduces itself, as well as the tags
of the libraries it uses.

We now revisit several of our definitions to account for \emph{tags} and
\emph{global specifications}.
A library interface now additionally holds the tags it introduces as well as those it uses. 
For instance, the $\Ltrans$ library described above depends on no tag and
introduces tags $\Ttag$ 
and $\perTag$. 
%
\begin{definition}[Interfaces]\label{def:library-interface-v2}
  An \emph{interface} is a tuple \(\genlibint = \tuple{\methods, \methodsconstr,
  \loc, \allowbreak\tagset, \allowbreak \tagsetdep}\), where $\methods$, $\methodsconstr$,
  and $\loc$ are as in \cref{def:library-interface-v1}, $\tagset$ is the set of
  tags $\genlibint$ introduces, and $\tagsetdep$ is the set of tags $\genlibint$ uses.
  The set of tags usable by~$\genlibint$ is $\tagsof(\genlibint) \eqdef
  \genlibint{.}\tagset \cup \genlibint{.}\tagsetdep$.
\end{definition}

We next define the notion of tagged method invocations (where a method invocation is associated with a set of tags). 
Hereafter, our notions of events, history (and so forth) use tagged method invocations (rather than methods invocations).  

\begin{definition}
  Given a library interface~$\genlibint$, a \emph{tagged method invocation} is
  of the form $\genmethlabthreadtag$, where the new component is a set of tags~$T
  \subseteq \tagsof(\genlibint)$.

\end{definition}

A \emph{global specification} of a library interface~$\genlibint$ is a set of
histories with some ``anonymized'' events. These are formalized using a designated library interface, $\libtags{\genlibint}$ (with a single method $\star$), which can
be tagged with any tag from~$\tagsof(\genlibint)$.
\begin{definition}
  Given an interface~$\genlibint$, the
  interface~$\libtags\genlibint$ is $\tuple{\{\star\}, \emptyset, \emptyset, \emptyset, \tagsof(\genlibint)}$.
\end{definition}
Now, given any history~$\hist \in \Hist{\{\genlibint\} \cup \coll}$, let $\anonymize{\genlibint}{\hist} \in \Hist{\{\genlibint, 
\libtags\genlibint\}}$ denote the \emph{anonymization} of $\hist$ such that each non-$\genlib$ event $e$ in $\hist$ labelled with a
method~$\genmethlabthreadtag$ of $\genlibint' \in \coll$ is replaced with
$\star^\gentags_\thread$ of $\libtags\genlibint$ if~$\gentags \neq \emptyset$
and is discarded otherwise.
It is then straightforward to extend the notion of libraries with global specifications as follows.
\begin{definition}\label{def:library-specification-v2}
  A \emph{library specification} $\genlib$ is a tuple $\tuple{\genlibint,
  \libdeps, \libConsistent, \libWellformed, \globalConsistent,
  \globalWellformed}$, where 
  $\genlibint$, $\libConsistent$ and $\libWellformed$ are as in \cref{def:library-specification-v1};
  $\globalConsistent$ and $\globalWellformed \subseteq \Hist{\{\genlibint, \libtags\genlibint\}}$ are the \emph{globally consistent} and \emph{globally well-formed} histories, respectively; 
  and $\libdeps$ denotes the \emph{tag-dependencies}, \ie a collection of libraries that provide all tags that $\genlib$~uses: $\genlibint.\tagsetdep \subseteq
  \bigcup_{\genlib' \in \libdeps} \genlib'.\tagset$.
  Both $\globalWellformed$ and~$\globalConsistent$ contain the empty history.
\end{definition}

In the context of a history, we write~$\makeset{\Ttag}$ for the set of events or calls
tagged with the tag~$\Ttag$ (we consider a return event tagged the same way as
its unique matching invocation).
%
%

For the $\Ltrans$ library, the \emph{globally} well-formed
set~$\Ltrans{.}\globalWellformed$ comprises histories~$\hist$ such that
for each thread $\thread$, $E[\thread]$ restricted to \textsf{PTBegin}, \textsf{PTEnd} and events of the form
$\Ttag$-tagged events is of the
form described by the regular expression $(\textsf{PTBegin}.\makeset{\Ttag}\!{}^*. \textsf{PTEnd})^*$.
In particular, transaction nesting is disallowed in our simple $\Ltrans$ library.

To define global consistency, we need to know when two operations are part of
the same transaction. Given a history~$\hist$, we define the \emph{same-transaction} relation, $\ksamet$,
relating pairs of $e, e' \in \makeset{\Ttag} \cup \textsf{PTEnd} \cup
\textsf{PTBegin}$ executed by the same thread $\thread$ such that
there is no $\textsf{PTBegin}$ or $\textsf{PTEnd}$ executed by $\thread$ between
them.
The set $\Ltrans{.}\globalConsistent$ of globally consistent histories contains
histories $\hist$ such that $\forall (e, e') \in \ksamet, e \in
\makeset{\perTag} \Leftrightarrow e' \in \makeset{\perTag}$, and all
completed $\textsf{PTEnd}$ calls are in $\makeset{\perTag}$.
Since the \textsf{PTEnd} call is related to all events inside its transaction,
this specification does express that (1) a transaction persist by the time the
call to \textsf{PTEnd} finishes and (2) all events persist \emph{atomically}.

Finally, we need to define the local consistency
predicate~$\Ltrans.\libConsistent$ describing the behavior of the registers
provided by~$\Ltrans$. This is where the we define the concrete meaning of `having
persisted' for these registers.
Let~$S$ be the sequential specification of a register. 
Let~$\hist \in \Hist{\Ltrans}$ be a history decomposed into $k$ eras
as $\hist_1 \cdot \Crash \cdot \hist_2 \cdot \Crash \cdot \cdots \Crash \cdot
\hist_k$.
Then~$\hist \in \Ltrans.\libConsistent$ iff all events are tagged with~$\Ttag$,
and there exists a $\happensbefore$-linearization~$\hist_\ell$ of
$
  \big(\left(\hist_1 \cdot \Crash \cdot \hist_2 \cdot \Crash \cdot \cdots \Crash
  \cdot \hist_{k-1}\right) \cap \makeset{\perTag} \big) \cdot \hist_k
$ 
such that~$\hist_\ell \in S$, where $\makeset{\perTag}$ is the set of events
of~$\hist$ tagged with~$\perTag$.
In other words, a write operation is seen after a crash iff it has persisted.
The requirement that such operations must appear within transactions and the
guarantee that they persist at the same time in a transaction are covered by the
global specifications.

\subsection{Library Implementations}
We have described how to \emph{specify} persistent libraries in our framework, and next describe how to \emph{implement} persistent libraries. 
This is formalized by the judgment $\libimpl{\coll}{I}{\genlib}$, stating that
\emph{$I$ is a correct implementation of library~$\genlib$ and only uses calls
in the collection of libraries $\coll$}.
As usual in such `layered' frameworks \citep{deep-specifications, yacovet},
the base layer, which represents the primitives of the hardware, is specified
as a library, keeping the framework uniform.
This judgement can be composed vertically as follows, where $\combineimpls{\genimpl}{\genimpl_L}$ denotes replacing the calls to library~$\genlib$ in $\genimpl$ with their implementations given by~$\genimpl_L$ (which in turn calls libraries~$\coll'$):
\begin{center}
$
  \prftree
  {\libimpl{\coll, \genlib}{\genimpl}{\genlib'}}
  {\libimpl{\coll'}{\genimpl_L}{\genlib}}
  {\libimpl{\coll, \coll'}{\combineimpls{\genimpl}{\genimpl_L}}{\genlib'}}
$
\end{center}
As we describe later,
this judgment denotes \emph{contextual refinement} and is impractical to prove directly.
We define a stronger notion that is \emph{compositional} and
more practical to use.

\begin{definition}[Implementation]
  Given a collection $\coll$ of libraries and a library~$\genlib$, an
  \emph{implementation~$\genimpl$ of~$\genlib$ over~$\coll$} is a
  map,
  $
    I \;\;:\;\; \genlib{.}\methods \times \Valbot \;\;\longrightarrow\;\; \Pow(\Hist\coll)
  $, 
  such that it is downward-closed:
  \begin{enumerate*}[label=\arabic*)]
  	\item  if~$\hist \in I(m(\vec v)_\thread, \valbot)$ and $\hist'$~is a prefix of~$\hist$, then~$\hist' \in I(m(\vec v), \bot)$;  and 
  	\item each $I(\genmethlabthread)$ history only contain events by thread~$\thread$.
\end{enumerate*}    
  
\end{definition}
%
Intuitively, $I(m(\vec v), \valbot)$ contains the histories corresponding to a call~$m(\vec v)$ with outcome~$\valbot$, where~$\valbot =
\bot$ denotes that the call has not terminated yet and $\valbot = \genval \in
\Val$ denotes the return value.
Downward-closure means that an implementation contains all partial histories.
We use a concrete programming language to write these
implementations; its syntax and semantics are standard and given in the
technical appendix.

For example, the implementation of $\Ltrans$ over $\Lweakreg$ and~$\Lqueue$
is given in \cref{fig:Ltrans-implementation}.
\begin{figure}[t]
  \begin{minipage}[t]{0.49\linewidth}
    \begin{lstlisting}[language=Imp]
    globals log := Q.new()
    method PTNewReg() := alloc(1)
    method PTRead(l) := read(l)
    method PTWrite(l, v) :=
      Q.append(log, (l, v));
      write(l, v)
    method PTBegin() := FENCE();
    method PTEnd() :=
      Append(log, COMMITTED);
      FENCE()
    \end{lstlisting}
  \end{minipage}
  \begin{minipage}[t]{0.49\linewidth}
    \begin{lstlisting}[language=Imp]
    method PTRecover() :=
      let w = Q.new() in
      while (x := Q.pop(log))
        if (x = COMMITTED)
          w = Q.new();
        else
          Q.append(w, x);
      while ((l, v) = Q.pop(log)) {
        write(l, v); }
    \end{lstlisting}
  \end{minipage}
  \caption{Implementation of $\Ltrans$}
  \label{fig:Ltrans-implementation}
\end{figure}
The idea is to keep an undo-log as a persistent queue that tracks
the values of the variables \emph{before} the transaction begins.
At the end of a transaction, and after all its writes have persisted, we
write the sentinel value \lstinline|COMMITTED| to the log to indicate that the
transaction was completed successfully. 
After a crash, the recovery routine \lstinline|PTRecover| returns the undo-log
and undoes the operations of \emph{incomplete} transactions by writing their
previous values.

\paragraph{Histories and Implementations}
An implementation~$I$ of~$\genlib$ over~$\coll$ is correct if
for all histories~$\hist \in \Hist{\{\genlib\} \cup \coll'}$ that use
library~$\genlib$ as well as those in~$\coll'$, and all
histories~$\hist'$ obtained by replacing calls to $\genlib$ methods with their implementation in $I$, if $\hist'$ is consistent,
then so is~$\hist$ (it satisfies the~$\genlib$ specification).

We define the action~$\hist \bind \genimpl$ of an implementation~$I$ on an abstract history~$\hist$ in a `relational' way: $\hist' \in \hist \bind \genimpl$ when we can \emph{match} each operation~$m'(\vec v)$ in~$\hist'$ with some operation $f(m'(\vec v))$ in~$\hist$ in such a way that the collection~$\finv{f}(\genmethlabthread)$ of
operations corresponding to some call~$\genmethlabthread$ in $\hist$ agrees
with~$\genimpl(\genmethlabthread)$.
\begin{definition}
  Let $\genimpl$ be an implementation of~$\genlib$ over~$\coll$; let~$\hist
  \in \Hist{\{\genlib\} \cup \coll'}$ and~$\hist' \in \Hist{\coll \cup \coll'}$ be two
  histories.
  Given a map~$f: \{1, \ldots, \norm{\hist'}\} \to \{1, \ldots,
  \norm{\hist} \}$, \emph{$\hist'$ $(I,f)$-matches $\hist$} if the following hold:
  \begin{enumerate}
    \item $f$~is surjective; \label{item:match_surjective}
    \item for all invocations of~$\hist$, if $m(\vec v)_\thread \notin
      \genlib{.}\methods$, then $f(m(\vec v)_\thread) = m(\vec v)_\thread$;
      \label{item:match_notL}
    \item for all threads~$\thread$, if $e_1$ precedes $e_2$ in
      $\hist'[\thread]$, then $f(e_1)$ precedes~$f(e_2)$ in~$\hist[\thread]$;
      \label{item:match_hb}
    \item for all calls~$\genmethlabthread$ of~$\hist$, the
      set~$\finv{f}(m(\vec v)_\thread)$ corresponds to a substring~$\hist'_m$
      of~$\hist'[\thread]$ and~$\hist'_m \in \genimpl(\genmethlabthread)$, where
      $\valbot$ is the (optional) return value of~$m(\vec v)_\thread$ in~$\hist$.
      \label{item:match_calls}
  \end{enumerate}
  The \emph{action} of~$\genimpl$ on a history~$\hist$ is defined as
 $
    \hist\bind\genimpl \;\coloneqq\; \{ \hist' \mid \exsts{f}{\text{$\hist'$
    $(I,f)$-matches $\hist$}} \}.
  $
\end{definition}
Condition \ref{item:match_surjective} ensures that all events of the abstract history are matched with an implementation event; 
condition~\ref{item:match_notL} ensures that the events that
do not belong to the library being implemented ($\genlib$) are left untouched, and
condition~\ref{item:match_hb} ensures that the thread-local order of events in the
implementation agrees with the one in the specification.
The last condition (\ref{item:match_calls}) states that the events corresponding to the
implementation of a call~$m(\vec v)$ are consecutive in the history of
the executing thread $\thread$, and correspond to the implementation~$\genimpl$.

\paragraph{Well-formedness and Consistency}
Recall that libraries specify both how they should be used (\emph{well-formedness}), and what they guarantee if used correctly (\emph{consistency}).
Using these specifications (expressed as sets of histories) to define implementation correctness is more subtle than one might expect.
Specifically, if we view a program using a library~$\genlib$ as a
downward-closed set of histories in~$\Hist{\genlib}$, we cannot assume all
its histories are in the set~$\genlib{.}\libWellformed$ of well-formed
histories, as the semantics of the program will contain \emph{unreachable}
traces (see~\citep{yacovet}).
To formalize reachability at a semantic level, we define \emph{hereditary
consistency}, stating that each step in the history was consistent, and
thus the current `state' is reachable.
\begin{definition}[Consistency]
  History~$\hist \!\in\! \Hist\coll$ is \emph{consistent} if for all~$\genlib
  \!\in\! \coll$, $\hist[\genlib] \!\in\! \genlib{.}\libConsistent$ and
  $\anonymize{\genlib}{\hist} \!\in\! \genlib{.}\globalConsistent$.
  It is \emph{hereditarily consistent} if all~$\hist[1..k]$ are consistent,
  for~$k \leq \norm{\hist}$.
\end{definition}
This definition uses the `anonymization' operator~$\anonymizeOp{\genlib}$
defined in \cref{sec:tags-and-global-specs} to test that the history~$\hist$
follows the global consistency predicates of every~$\genlib \in \coll$.

We further require that programs using libraries respect \emph{encapsulation}, defined below, 
stating that locations obtained from a library constructor are only used
by that library instance.
Specifically, the first condition ensures that distinct constructor calls return distinct locations. 
The second condition ensures that a non-constructor call $e$ of $\genlib$ uses locations that have been allocated by an earlier call $c$ ($c \happensbefore e$) to an $\genlib$ constructor.
\begin{definition}[Encapsulation]\label{def:encapsulation}
  A history $\hist \in \Hist\coll$ is \emph{encapsulated} if the following hold, where  $C$ denotes the set of calls to constructors in~$\hist$:
  \begin{enumerate}
    \item for all~$c, c' \in C$, if $c \neq c'$, then $\loc(c) \cap \loc(c') =
            \emptyset$;
    \item for all $e \in \hist \setminus C$, if $\loc(e) \neq \emptyset$, then
      there exist $c \in C$, $\genlib \in \coll$ such that $e, c \in
      \genlib.\methods$, $c \happensbefore e$ and $\loc(e) \suq \loc(c)$.
  \end{enumerate}
\end{definition}

We can now define when a history of~$\coll$ is
\emph{immediately well-formed}: it must be encapsulated and be
well-formed according to each library in~$\coll$ and all the
tags it uses.

\begin{definition}
History~$\hist \in \Hist{\coll}$ is \emph{immediately well-formed} if the following hold:
\begin{enumerate}
  \item $\hist$ is encapsulated;
  \item $\hist[\genlib] \in \genlib{.}\libWellformed$, for all~$\genlib \in
    \coll$; and
  \item $\anonymize{\genlib}{\hist} \in \genlib{.}\globalWellformed$ for
    all $\genlib \in \TagDep(\coll)$, where the immediate
    dependencies~$\TagDep(\coll)$ is defined as $\bigcup_{\genlib \in
  \coll} \{ \genlib \} \cup \libdeps(\genlib)$. 
\end{enumerate}
\end{definition}
We finally have the notions required to define a \emph{correct implementation}.

\paragraph{Implementation Correctness}
As usual, an implementation is correct if all behaviors of the implementation
are allowed by the specification. In our setting, this means that if a
concrete history is \emph{hereditarily consistent}, so should the abstract
history. Moreover, assuming the abstract history is well-formed, all
corresponding concrete histories should also be well-formed; this corresponds
to the requirement that the library implementation uses its dependencies
correctly, under the assumption that the program itself uses its libraries
correctly.

\begin{definition}[Correct implementation]
\label{def:sc_global_correctness}
  An implementation $\genimpl$ of~$\genlib$ over~$\coll$ is \emph{correct}, written
  $\libimpl{\coll}{\genimpl}{\genlib}$, if for all collections~$\coll'$, all `abstract' histories~$\hist \in \Hist{\{\genlib\} \cup \coll'}$ and
 all `concrete' histories~$\hist' \in \hist \bind \genimpl \subseteq
  \Hist{\coll \cup \coll'}$, the following hold:
  \begin{enumerate}
    \item if $\hist$ is immediately well-formed, then $\hist'$ is also immediately well-formed; and
    \item if $\hist'$ is immediately well-formed and hereditarily consistent, then $\hist$ is consistent.
  \end{enumerate}
\end{definition}
This definition is similar to \emph{contextual refinement} in that it quantifies over all contexts: it considers histories that use arbitrary libraries as well as those that concern~$\genimpl$ directly.
We now present a more convenient, \emph{compositional} method for proving an implementation correct, which allows one to only consider libraries and tags that are used by the implemented library.

\subsection{Compositionally Proving Implementation Correctness}
\label{subsec:sc_composition}
Recall that in this section we present our framework in a simplified
sequentially consistent setting; later in \cref{sec:framework} we generalize our framework to the weak memory setting.
%
We introduce the notion of \emph{compositional correctness}, simplifying the global correctness conditions in \cref{def:sc_global_correctness}. 
Specifically, while \cref{def:sc_global_correctness} considers histories with arbitrary libraries that may use tags introduced by~$\genlib$, our compositional condition requires one to prove that only those $\genlib$ methods that are $\genlib$-tagged satisfy $\genlib.\globalConsistent$.

\begin{definition}[Compositional correctness]
\label{def:sc_compositional_correctness}
  An implementation $\genimpl$ of~$\genlib$ over~$\coll$ is
  \emph{compositionally correct} if the following hold:
  \begin{enumerate}
    \item \label{item:sc_compositional_correctness_wf}For all $\coll'$, $\hist \in \Hist{\{\genlib \} \cup \coll}$ and
      $\hist' \in \hist \bind \genimpl \subseteq \Hist{\coll \cup \coll'}$, if
      $\hist'$ is well-formed, then $\hist$ is well-formed;
    \item \label{item:sc_compositional_correctness_consistent1}For all $\hist \in \Hist{\genlib}$ and $\hist' \in \hist \bind
      \genimpl \subseteq \Hist{\coll}$, if $\hist'$ is well-formed
      and hereditarily consistent, then $\hist \in \genlib.\libConsistent \cap
      \genlib.\globalConsistent$; and
    \item \label{item:sc_compositional_correctness_consistent2}For all~$\genlib' \in \coll$, $\hist \in \Hist{\{\genlib, \genlib',
      \libtags{\genlib'}\}}$ and $\hist' \in \hist \bind \genimpl$, if
      $\anonymize{\genlib'}{\hist'} \in \genlib'.\globalWellformed \cap
      \genlib'.\globalConsistent$, then $\anonymize{\genlib'}{\hist} \in
      \genlib'.\globalConsistent$.
  \end{enumerate}
\end{definition}
The preservation of well-formedness (condition \ref{item:sc_compositional_correctness_wf}) does not change compared to its counterpart in \cref{def:sc_global_correctness}, as
in practice this condition is easy to prove directly.
Condition \ref{item:sc_compositional_correctness_consistent1} requires one to prove that the implementation is correct \emph{in isolation} (without $\coll'$).
Condition \ref{item:sc_compositional_correctness_consistent2} requires one to prove that global consistency requirements are maintained for all dependencies of the implementation.
In practice, this corresponds to proving that those $\genlib$ operations tagged with existing tags in~$\coll$ obey the global
specifications associated with these tags.
Intuitively, the onus is on the library that \emph{uses} a tag for its methods
to prove the associated global consistency predicate: we need not consider
unknown methods tagged with tags in~$\genlib.\tagset$.

Finally, we show that it is sufficient to show an implementation $\genimpl$ is
compositionally correct as it implies that $\genimpl$ is correct. 
\begin{theorem}[Correctness]
  If an implementation $\genimpl$ of~$\genlib$ over $\coll$ is
  \emph{compositionally correct} (\cref{def:sc_compositional_correctness}), then
  it is also \emph{correct} (\cref{def:sc_global_correctness}).
\end{theorem}

\begin{example}[Transactional Library $\Ltrans$]
Consider the implementation~$\implTrans$ of~$\Ltrans$ over~$\coll = \{\Lweakreg, \Lqueue\}$ given in \cref{fig:Ltrans-implementation}, and let us assume we were to show that $\implTrans$ is compositionally correct. 
Our aim here is only to outline the proof obligations that must be discharged; 
later in \cref{sec:transactions} we give a full proof in the more general weak memory setting.
\begin{enumerate}
  \item For the first condition of compositional correctness, we must
    show~$\implTrans$ preserves well-formedness:  if the abstract
    history~$\hist$ is well-formed, then so is any corresponding concrete
    history~$\hist' \in \hist \bind \implTrans$.
    This is straightforward as the well-formedness conditions of $\Lweakreg$ and
    $\Lqueue$ are trivial, and~$\Ltrans$ does not use any existing tag.
  \item For the second condition of compositional correctness, we must show that
    $\implTrans$ preserves consistency in the other direction: keeping the
    notations as above, assuming $\hist'$ is consistent for~$\Lambda$, then
    $\hist$ is consistent as specified by~$\Ltrans$. There are two parts to this
    obligation, as we also have to show that the $\Ltrans$'s operations tagged
    with $\Ttag$ satisfy the global consistency predicate of the library.
  \item The last condition holds vacuously as~$\Ltrans$ does not use any existing
    tags.
\end{enumerate}
\end{example}

\begin{example}[A Client of $\Ltrans$]
To see how the global consistency specifications work, consider a simple min-max counter library, $\Lmmcounter$, tracking the maximal and minimal integer it has been given. 
The $\Lmmcounter$ is to be used within $\Ltrans$~transactions, and provides four methods: $\mmcounterNew()$ to construct a min-max counter, 
$\mmcounterAdd(x, n)$, to add integer~$n$ to the min-max counter, and
$\mmcounterMin(x)$ and~$\mmcounterMax(x)$ to read the respective values.

We present the $\implmmcounter$ implementation over~$\Ltrans$ in  \cref{fig:Lmmcounter-implementation}.
\begin{figure}[t]
  \begin{minipage}[t]{0.49\linewidth}
    \begin{lstlisting}[language=Imp]
      method mmNew() := 
       (PTNewReg(), PTNewReg())

      method mmAdd(x, n) :=
        PTWrite(min(n, PTRead(x.1)))
        PTWrite(max(n, PTRead(x.2)))
    \end{lstlisting}
  \end{minipage}
  \begin{minipage}[t]{0.49\linewidth}
    \begin{lstlisting}[language=Imp]
      method mmMin(x) :=
        PTRead(x.1)

      method mmMax(x) :=
        PTRead(x.2)
    \end{lstlisting}
  \end{minipage}
  \caption{Implementation~$\implmmcounter$ of $\Lmmcounter$}
  \label{fig:Lmmcounter-implementation}
\end{figure}
The idea is simply to track two integers denoting the minimal
and maximal values of the numbers that have been added.
Interestingly, even though they are stored in $\Ltrans$
registers, the implementation does not begin or end transactions: this is the
responsibility of the client to avoid nesting transactions.
This is enforced by~$\Lmmcounter$ using a global well-formedness predicate.
Moreover, the $\mmcounterAdd$ operation is tagged with~$\Ttag$ from
the~$\Ltrans$ library, ensuring that it behaves well \wrt transactions.
A non-example is a version of~$\implmmcounter$ where the minimum is in a
$\Ltrans$~register, but the max is in a ``normal'' $\Lweakreg$ register. This
breaks the atomicity guarantee of transactions.

Formally, the interface~$\genlibint_\MMCounter$ has four methods
as above, where~$\mmcounterNew$ is the only constructor.
The set of used tags is~$\tagsetdep = \{ \Ttag, \perTag \}$, and all $\genlibint_\MMCounter$ methods are tagged with~$\Ttag$ as they all use primitives from~$\Ltrans$.
The consistency predicate is defined using the obvious sequential
specification~$\Smmcounter$, which states that calls to~$\mmcounterMin$
return the minimum of all integers previously given to~$\mmcounterAdd$ in the
sequential history.
We lift this to (concurrent) histories as follows.
A history~$\hist \in \Hist{\Lmmcounter}$ is in~$\Lmmcounter.\libConsistent$ if
there exists~$E_\ell \in \Smmcounter$ that is a $\happensbefore$-linearization
of $E_1[\perTag] \cdot E_2[\perTag] \cdots E_{n-1} \cdot E_n[\perTag]$, where $\hist$ constructs $n$ eras decomposed as $\hist = E_1 \cdot \Crash \cdots \Crash
\cdot E_n$ (recall
that~$E[\perTag]$ denotes the sub-history with events tagged with~$\perTag$,
that is, persisted events.).
The global specification and well-formedness conditions of~$\Lmmcounter$ are
trivial.
Because $\Lmmcounter$ uses tag $\Ttag$ of~$\Ltrans$, a well-formed history
of~$\Lmmcounter$ must satisfy $\Ltrans.\globalWellformed$, which requires
that all operations tagged with~$\Ttag$ be inside transactions, and
$\Ltrans.\globalConsistent$ guarantees that $\Lmmcounter$ operations persist
atomically in a transaction.

When proving that the implementation in \cref{fig:Lmmcounter-implementation} satisfies~$\Lmmcounter$ using
compositional correctness, one proof obligation is to show that, given
histories $\hist \in \Hist{\{\Ltrans, \Lmmcounter, \libtags{\Ltrans}\}}$ and $\hist' \in
\hist \cdot \genimpl_\MMCounter \subseteq \Hist{\{\Ltrans, \libtags{\Ltrans}\}}$, if
$\anonymize{\Ltrans}{\hist'} \in \Ltrans.\globalConsistent$, then
$\anonymize{\Ltrans}{\hist} \in \Ltrans.\globalConsistent$.
This corresponds precisely to the fact that min-max counter operations persist
atomically in a transaction, assuming the primitives it uses do as well.
\end{example}

\subsection{Generic Durable Persistency Theorems}
We consider another family of libraries with persistent reads/writes guaranteeing the following:
\begin{addmargin}[2em]{2em}
	if one replaces regular (volatile) reads/writes in a \emph{linearizable} implementation with persistent ones, then the implementation obtained is \emph{durably linearizable}.
\end{addmargin}
We consider two such such libraries: \Flit~\citep{Flit} and \Mirror~\citep{Mirror}.
Thanks to our framework, we formalise the statement above for the first time and prove it for both Flit and Mirror against a realistic consistency (concurrency) model (see \cref{sec:flit_and_mirror}).


\section{A General Framework for Persistency and Consistency}
\label{sec:framework}
We generalise our persistency specification framework from \cref{sec:overview} to account for an arbitrary (potentially weak) memory consistency (concurrency) model (rather than sequential consistency (SC) which was hard-coded into our formalism in \cref{sec:overview}). 
As such, we need to revisit and generalise several of our definitions.

\subsection{Plain Executions and Executions}

\paragraph{Plain Executions}
Unlike in the SC setting of \cref{sec:overview} where we represented an
execution as a totally-ordered history (sequence) of events, in the general weak
consistency setting, such a total execution order does not exist in general. 
As such, we model an execution as a pomset (partially ordered multiset) of events. 

\begin{definition}\label{def:pomset}
  A \emph{pomset} over the set~$X$ is a tuple $((E, \leq), \lambda)$ consisting
  of
  an ordered set~$E$ and a map~$\lambda: E \to X$.
  We write~$\Pomset(X)$ for the set of non-empty pomsets over~$X$.
  Two pomsets $((E, \leq), \lambda)$ and $((F, \leq), \mu)$ over~$X$ are
  identified if there exists an order-isomorphism~$f : E \to F$ such that $\mu
  \circ f = \lambda$.
  The underlying set~$E$ of a pomset $P = ((E, \leq), \lambda)$ is denoted
  by~$\toplain{P}$.
  %
\end{definition}

Following the literature on weak consistency models where the execution of each instruction is modelled by a \emph{single} event, we handle method calls differently from \cref{sec:overview}: rather than having two distinct events for each method invocation and return, we model each method call as a single event~$\genmethlab$, which is \emph{incomplete} if~$\valbot = \bot$ and
\emph{complete} otherwise. 
As such, an \emph{event} is either such a method call, or it is a crash event.
%
Given a library interface $\genlibint$ (as given in
\cref{def:library-interface-v2}), we can then model a \emph{plain execution} of
$\genlibint$ as a pomset $((E, \kpo), \lab)$, where $E$ is a set of
$\genlibint$ events, $\kpo$ is the \emph{program order}, and $\lab$ is the
\emph{label function}, associating each event with its label of the form
$\genmethlabtag$ or $\Crash$. Moreover, incomplete method calls are maximal
events in~$\kpo$ unless their immediate successor is a crash event.

\begin{definition}[Plain executions]
  \label{def:plain-execution}
  A \emph{plain execution}~$G$ of an interface~$\genlibint$ is a pomset $((E, \kpo), \lambda)$, where $E$ is a set of events, $\kpo$ is the program order and $\lambda: E \rightarrow \genlibint.\methods \cup \{\Crash\}$, such that the $\kpo$-immediate successor of an incomplete method call~$\methlab{m}{\vec v}{\bot}{}$ is a crash event.
 
  Given two plain executions, $G_1 = ((E_1, \kpo_1), \lambda_1)$ and $G_2 =
  ((E_2, \kpo_2), \lambda_2)$, their \emph{sequential composition} is $G_1 ; G_2
  \eqdef ((\tilde E, \tilde\kpo), \tilde\lambda)$, where $\tilde E \eqdef E_1
  \amalg E_2$ is the disjoint sum\footnotemark of events, $\tilde\lambda$ is the
  induced labelling, and $\tilde\kpo$ is the transitive closure of the following relation: 
  \[
    \kpo_1 \cup \kpo_2 \cup \setcomp{(e_1, e_2) \in E_1 \times E_2}{\text{$e_1$
    is labeled by a complete operation or $e_2 = \Crash$}}
  \]
  The set of plain executions is~$\PExec{\genlibint}$; for brevity, $((E, \kpo), \lab)$ is often written as~$\tuple{E,\kpo}$.
  \footnotetext{For example, define $X \amalg Y \coloneqq \{0\} \times X \cup
  \{1\} \times Y$.}
\end{definition}
Note that the sequential composition of two plain executions preserves the invariant that the only possible immediate $\kpo$-successors of an incomplete method call is a crash event.

\paragraph{Executions}
Recall from \cref{sec:overview} that a method call $C_1$ happens before another $C_2$ in a history $\hist$ ($C_1 \happensbefore_\hist C_2$) if the response of $e_1$ precedes the invocation of $C_2$ in the totally-ordered history $\hist$. 
This captures the real-time ordering present under the strong SC model. 
However, under weaker consistency models (\eg in modern multi-core processors) this notion of real-time ordering is not realistic. 
Indeed, the happens-before notion varies from one weak model to another, and is typically defined in terms of a \emph{synchronizes-with} relation, which itself is also model-dependent. 
As such, we record the happens-before and synchronizes-with relations as primitive notions within the definition of a library execution. 

Note that executions additionally track the \emph{tags} associated with method calls: events are labelled with \emph{tagged}
method calls~$\genmethlabtag$ as well as crash events.
Indeed, in general the tags are not observable by programs and belong to
executions: \eg the tag~$\Ptag$ (denoting that a write has persisted)
pertains to phenomena that are external to the program.

\begin{definition}[Library executions]
  \label{def:library-execution}
  Given a library interface $\genlibint$, a \emph{library $\genlibint$
  execution} is a tuple 
  $
    \genexec = \tuple{E, \kpo, \ksw, \klhb}
  $ such that:
  \begin{itemize}
    \item $\tuple{E, \kpo}$ is a pomset labeled with tagged events of~$\genlibint$;
    \item $\ksw \suq E \times E$ is the \emph{synchronizes-with} relation;
    \item $\klhb$ is the \emph{happens-before} relation, which is a strict order
      with $\kpo \cup \ksw \suq \klhb$.
  \end{itemize}
  The set of library $\genlibint$ executions is denoted $\Exec{\genlibint}$.
  Given a library execution $\genexec = \tuple{E, \kpo, \ksw, \klhb}$, its
  underlying plain execution $\tuple{E, \kpo}$ is denoted by
  $\toplain{\genexec}$. An execution~$\genexec$ \emph{refines} a plain
  execution~$G$, written $\genexec \refines G$, when $\toplain{\genexec} = G$.
  This definition is lifted to a collections $\coll$ of libraries by allowing
  events to be labelled with any library in $\coll$.
\end{definition}
We often use the `$\genexec.$' prefix to project the components of $\genexec$, \eg $\genexec.\ksw$.

\subsection{Library Specifications}
Most of our definitions pertaining library specifications remain unchanged from \cref{sec:overview}, and the only definition we need to adapt is the anonymization
operation~$\anonymizeOp{\genlibint} : \Exec{\{\genlib\} \cup \coll} \to \Exec{\{\genlib\} \cup 
\libtags{\genlibint}}$ which now operates on decorated pomsets.
To do this, we first change the execution labelling:
$\tilde{\genexec}$ is the same as~$\genexec$, except that the labelling
map, $\lab$, of the underlying plain execution is replaced with~$f \circ \lab$,
where the map~$f$ over events is defined as:
\begin{center}
$
  f(\ell) =
  \begin{cases*}
    \star^T & if $\ell = \genmethlabtag \notin \genlib.\methods$ \\
    \ell    & if $\ell \in \genlib.\methods$ or $\ell = \Crash$
  \end{cases*}
$
\end{center}
We then define $\anonymize{\genlibint}{\genexec}$ as the restriction
of~$\tilde\genexec$ to events which are not tagged with~$\emptyset$.

As in \cref{def:library-specification-v2}, a \emph{library specification} is a
tuple~$\tuple{\genlibint, \libdeps, \libConsistent, \libWellformed,
\globalConsistent, \globalWellformed}$, where $\genlibint$ and $\libdeps,$ are as before, and $ \libConsistent$, $\libWellformed$, $\globalConsistent$ and $\globalWellformed$ are sets of \emph{executions} (not histories) as in \cref{def:library-execution}.

\newcommand{\LReads}{\ensuremath{\mathcal{R}}}%
\newcommand{\LWrites}{\ensuremath{\mathcal{W}}}%
\newcommand{\LUpdates}{\ensuremath{\mathcal{U}}}%
\newcommand{\LFlushes}{\ensuremath{\mathcal{F\!\!L}}}%
\newcommand{\LFOpts}{\ensuremath{\mathcal{F\!O}}}%
\newcommand{\LMFences}{\ensuremath{\mathcal{MF}}}%
\newcommand{\LSFences}{\ensuremath{\mathcal{SF}}}%
\newcommand{\Reads}{\ensuremath{\mathit{R}}}%
\newcommand{\Writes}{\ensuremath{\mathit{W}}}%
\newcommand{\Updates}{\ensuremath{\mathit{U}}}%
\newcommand{\Flushes}{\ensuremath{\mathit{FL}}}%
\newcommand{\FOpts}{\ensuremath{\mathit{FO}}}%
\newcommand{\MFences}{\ensuremath{\mathit{MF}}}%
\newcommand{\SFences}{\ensuremath{\mathit{SF}}}%
\newcommand{\eklhb}{\ensuremath{\klhb_{\mathsf{e}}}}%
\begin{example}[\pxes]
  Our main example of a library with weak consistency and persistency is~$\pxes$ (the Intel-x86 consistency and persistency model \cite{Px86}).
  Our specification below is a simple adaptation of~\citep{Px86}.
  Let us begin with the library interface~$\genlibint_\pxes$, introducing
  two new tags: $\Dtag$, denoting events that are
  \emph{durable} in that they \emph{can} persist; and $\Ptag$, denoting events
  that did persist.
  The \emph{\pxes interface} is $\tuple{\methods, \methodsconstr, \loc,
  \{\Dtag, \Ptag\}, \emptyset}$, where:
  \begin{itemize}
    \item $\methodsconstr \eqdef \{\mathsf{alloc}()\}$;
    \item $\methodsdurable \eqdef \methodsconstr \cup \bigcup_{x \in \Loc}
            \methodsdurable^x$, where $\methodsdurable^x \eqdef \LWrites^x \cup
            \LUpdates^x \cup \LFlushes^x \cup \LFOpts^x$; and:\\
          $\LWrites^x \eqdef \setcomp{\store{x}{v}}{v \!\in\! \Val}$ is the set of write events on location $x$; \\
          $\LUpdates^x \eqdef \setcomp{\mathsf{Upd}(x, v, v')}{v, v'\! \!\in\! \Val}$ is the set of read-modify-write operations on location $x$;\\
          $\LFlushes^x \eqdef \{ \flushC x\}$ is the set of synchronous flush events on location $x$; and\\
          $\LFOpts^x \eqdef \{ \foC x \}$ is the set of delayed flush (flush-opt) events on location $x$.

    \item $\methods \eqdef \methodsconstr \cup \methodsdurable \cup \LMFences \cup \LSFences \cup \bigcup_{x \in \Loc} \LReads^x$, where
          $\LMFences {\eqdef} \{\mfenceC\}$ is a memory fence invocation,
          $\LSFences {\eqdef} \{\sfenceC\}$ is a store fence, and
          $\LReads^x {\eqdef} \{\load x\}$ is a read from location $x$.
    \item $\for{x \in \Loc} \for{i \in \methodsdurable^x \cup \LReads^x} \loc(i, \valbot) \eqdef \{x\}$, $\for{x}\loc(\mathsf{alloc}(), x) \eqdef \{ x \}$ and otherwise
         $\loc(l, \valbot) \eqdef \emptyset$.
  \end{itemize}
  
  Given an execution $\genexec \in \Exec{\LIpxes}$ with $\genexec = \tuple{E, \kpo,
  \ksw, \khb}$, 
%
  let $\Reads \eqdef \bigcup_{x \in \Loc} \Reads^x$ with $\Reads^x \eqdef \setcomp{e \in E}{\lab(e) \in \LReads^x}$, and let $\Writes^x$, $\Writes$,
  $\Updates^x$, $\Updates$, $\Flushes^x$, $\Flushes$, $\FOpts^x$, $\FOpts$,
  $\MFences$ and $\SFences$ be defined analogously. Let us also define the following sets: 
  \begin{align*}
    D & \eqdef \setcomp{e \!\in\! E}{ \lab(e) \in \methodsdurable } && \text{durable events}\\
    D^x &= \setcomp{e \!\in\! D}{\loc(e) {=} \{x\}}  && \text{durable events on location }x\\
    \eklhb &= \klhb \cap \setcomp{(a, b)}{(a, b) \notin \kpo \cup \finv\kpo}  && \khb\text{ between different threads}\\
    \keb &= \kpo ; \Crash ; \kpo  && \text{the `era-before' relation}\\
    \kse &= \{ (e, e') \mid (e, e') \notin \keb \cup \finv{\keb} \}  && \text{the `same-era' relation}
  \end{align*} 
  Given a relation $r$ on $E$, let $\intE r \eqdef r \cap \kse$
  Execution $\genexec$ is \emph{\pxes-consistent} if there exists a
  \emph{reads-from} relation~$\krf \subseteq \Writes \times \Reads$ such that $\finv\krf$ is total and functional (\ie every read is related to exactly one write), a strict order~$\ktso$
  that is total on~$W \cup U$ and a strict order~$\knvo$ on~$D$ such that
  $\krf, \ktso, \knvo \subseteq \keb \cup \kse$
  and:
  \newcommand{\vsep}{\\[-2pt]}%
          \begin{align}
             & \klhb \cup \tso \text{ is acyclic and	}\kcom \subseteq \tsoSE \cup \kpo
            \tag{\textsc{A1}} \label{ax:common1}  \vsep
             & \for{x \!\in\! \Loc} \for{(w, r) \!\in\! \kcom_x} \for{w' {\in} \Writes^x\! \cup \!\Updates^x}
            (w'\!, r) \!\in\! \tsoSE \cup \kpo \Rightarrow (w, w') \!\not\in\! \tsoSE
            \tag{\textsc{A2}} \label{ax:common2} \vsep
             & ([E]; \poSE;  [\MFences \cup \Updates]) \cup ([\MFences \cup \Updates \cup \Reads]; \poSE; [E]) \subseteq \tsoSE  \tag{\textsc{A3}} \label{ax:common3} \vsep
             & ([E]; \poSE;  [\SFences]) \cup ([\SFences]; \poSE; [E \setminus \Reads]) \subseteq \tsoSE  \tag{\textsc{A4}} \label{ax:common4} \vsep
             & [\Writes \cup \Flushes]; \poSE; [\Writes \cup \Flushes] \suq \tsoSE \tag{\textsc{A5}} \label{ax:common5} \vsep
             & \for{x \in \Loc} ([\Flushes^x]; \poSE; [\FOpts^x]) \cup ([\FOpts^x]; \poSE; [\Flushes^x]) \cup ([\Writes^x]; \poSE; [\FOpts^x]) \suq \tsoSE  \tag{\textsc{A6}} \label{ax:common6} \vsep
             & \for{x \in \Loc}\rest{\tsoSE}{D^x} \suq \nvoSE \tag{\textsc{A7}} \label{ax:common7}\vsep
             & \for{x \in \Loc} [D^x ]; \intE{(\tso \cup \eklhb)}; [\FOpts^x  \cup \Flushes^x ] \subseteq \knvo  \tag{\textsc{A8}} \label{ax:common8} \vsep
             & ([\Flushes]; \tsoSE; [D]) \cup ([\FOpts]; \poSE; [\MFences \cup
             \SFences \cup \Updates]; \tsoSE; [D]) \subseteq \nvoSE
              \tag{\textsc{A9}} \label{ax:common9} \vsep
             & \for{(w, r) \in \krf \cap \keb} w \in
             \makeset{\Ptag} \;\land\; (\diagonal{\{w\}} ; \knvo ;
             \diagonal{\makeset{\Ptag} \cap W^{\loc(w)}}; \keb;
             \diagonal{\{r\}}) = \emptyset   \tag{\textsc{New}} \label{ax:new}
          \end{align}
        \end{example}

The specification of $\pxes$ above is adapted from~\citep[Def.~4]{Px86}.
Specifically, axioms \eqref{ax:common1}--\eqref{ax:common9} are directly imported from \cite{Px86}. 
  The main difference is that instead of considering \emph{execution chains}, which are sequences of executions, we use our more general approach of a single execution with crash events. This generality is required for general data structures such as queues, as the pre-crash events cannot be summarized by a set of initial events, unlike in
  \pxes where the `maximal' write to each location captures the behavior of that
  location after a crash.
  This is reflected in \eqref{ax:new}, stating that a read $r$ that reads from a write $w$ in an earlier era, must do so from the maximal persisted such write on the same location, \ie there should be no intervening persisted writes on the same location between $w$ and $r$.
  Moreover, $\krf, \ktso, \knvo \subseteq \keb \cup \kse$ ensures that events in later eras are not $\krf$-, $\ktso$- or $\knvo$-related to those in earlier eras.
  
\subsection{Library Implementations}
We next describe how our general framework can be used to verify library implementations.
\subsubsection{Semantic implementations}
Analogously to \cref{sec:overview}, a semantic implementation in this general setting is a downward-closed map from method calls to sets of plain executions. 
This time, we define downward-closure with respect to the \emph{prefix order} over plain executions.
\begin{definition}
Given a collection $\coll$ and plain executions~$\genplainexec, \genplainexec' \in
  \PExec\coll$, the plain execution $\genplainexec$ is an \emph{immediate prefix} of $\genplainexec' \in \PExec\coll$, written $\genplainexec \prefimm
  \genplainexec'$, if there exists a $\kpo$-maximal event~$e \in \genplainexec'.E$ such
  that~$\genplainexec' \setminus \{ e \} = \genplainexec$.
  The \emph{prefix order}, $\pref$, is the transitive closure of~$\prefimm$.
  Both definitions are lifted naturally to library executions: 
  $\genexec\prefimm
  \genexec'$ iff $\toplain{\genexec} \prefimm \toplain{\genexec'}$, and $\pref$ on library executions is the transitive closure of $\prefimm$ on library executions. 
  Both prefix relations are defined on executions by considering~$\khb$-maximal
  events.
\end{definition}
Intuitively, $\genplainexec \prefimm \genplainexec'$ holds when $\genplainexec'$ can
be obtained by adding a new event $e$ to~$\genplainexec$, corresponding to
a single step of program execution by one thread.
When interpreting programs, we consider all their partial executions, and as
such, any immediate prefix of an execution of a program will also be in its
semantics.
We can now define \emph{library implementations} as indexed sets of executions that
satisfy this downward-closure property.
\begin{definition}
  An \emph{implementation} $I$ of library~$\genlib$ over collection~$\coll$ is a map, 
 $
    I : \genlib.\methods \times \Valbot
    \;\;\longrightarrow\;\; \PExec{\Lambda}
 $
  that is \emph{downward-closed}: for all non-empty executions~$G$ and~$G'$, if
  $G \pref G'$ and $G' \in I(\genmethlab)$, then $G \in I(\methlab{m}{\vec
  v}{\bot}{})$, where we identify a method call with the pair of the invocation
  and the return value.
\end{definition}

\subsubsection{A simple Concurrent Programming Language}
\begin{figure}[t]
  %
  \footnotesize
  \[
    \begin{array}[t]{@{} l @{\hspace{40pt}} l @{}}
      \begin{array}[t]{@{} l @{}}
        \textbf{Basic domains} \\
        \;\; \begin{array}{@{} l @{\hspace{10pt}}  r @{}}
               \a \in \Reg
                            & \text{Registers}    \\
               v \in \Val   & \text{Values}       \\
               \thread \in \ThreadIds
                            & \text{Thread IDs}   \\
               m \in \MName & \text{Method names}
             \end{array}
        %
      \end{array}
       &
      \begin{array}[t]{@{} l @{}}
        \textbf{Expressions, sequential commands and programs} \\
        \;\;\begin{array}{@{} r @{\hspace{2pt}} l @{}}
              \Exp \ni \pexp ::=
                                     & v \mid \a \mid \pexp + \pexp \mid \cdots                                         \\
              \Com \ni \com ::=
                                     & \skipC \mid \assign \a \pexp \mid \assign \a m(\b_1 \cdots \b_n) \mid \continueC \\
                                     & \mid  \com; \com \mid \iteC{\pexp}{\com}{\com} \mid \whileC \pexp \com           \\
                                     & \returnC{e} \\
              \prog \in \Prog \eqdef & \ThreadIds \fm \Com
            \end{array}
      \end{array}
    \end{array}
  \]
  %
  \vspace{-10pt}
  \caption{A simple concurrent programming language}
  \label{fig:language-syntax}
\end{figure}
In \cref{fig:language-syntax} we define a simple concurrent language for library implementations in our case studies (\cref{sec:transactions}, \cref{sec:flit_and_mirror}).
The semantics of a program~$\prog$, written $\sem{\prog}$, is defined as a set of plain executions and is standard (see \cite{yacovet}).
We write $\sem{\prog}^v$ for the set of plain executions where $\prog$ returns value~$v$, and write~$\sem{\prog}^\bot$ for the set of  plain executions where $\prog$ has not yet returned.
%

There are two contexts in which the programming language is used: to define
library implementations and to define a top-level program.
A \emph{syntactic} library implementation~$\gensynimpl$ of~$\genlib$ over~$\coll$ is a program~$\gensynimpl(m(\vec x))$ with variables~$\vec
x$ for each method~$m$ of~$\genlib$. 
This defines a \emph{semantic} implementation~$\sem{\gensynimpl}$ with~$\sem{\gensynimpl}(\genmethlab) \eqdef
\sem{\gensynimpl(m(\vec x))[\vec x \coloneqq \vec v]}^\valbot$, where$[\vec x \coloneqq \vec v]$ denotes the point-wise substitution of $\vec x$ with $\vec v$. 

The semantics~$\semprog{\prog}$ of a top-level program is different as we must
account for crashes. 
This is done by restarting the program from scratch after each crash:
\begin{center}
$
  \semprog{\prog}^\valbot \;=\;
  \setcomp
  {\genplainexec_1 \cdot \Crash \cdots \Crash \cdot \genplainexec_n}
  {\for{i < n} \genplainexec_i \in \sem{\prog}^\bot \text{ and } \genplainexec_n \in
  \sem{\prog}^\valbot}
$
\end{center}
Formally, an execution of~$\prog$ is any number of partial executions
interrupted by a crash, followed by a possibly complete execution.

\subsubsection{Semantic Substitutions}
We define the action of an implementation~$\genimpl$ of a library~$\genlib$ over~$\coll$ on a plain execution~$\genplainexec \in \PExec{\genlib}$, which
captures what happens when a program that calls~$\genlib$ is `linked'
with the $\genlib$ implementation~$\genimpl$.
This operation, defined at the level of pomsets, is crucial to define what it
means for the implementation to be correct.

\begin{figure}[t]
  \begin{minipage}{0.37\linewidth}
    \phantom{x}
    \includegraphics[width=\linewidth]{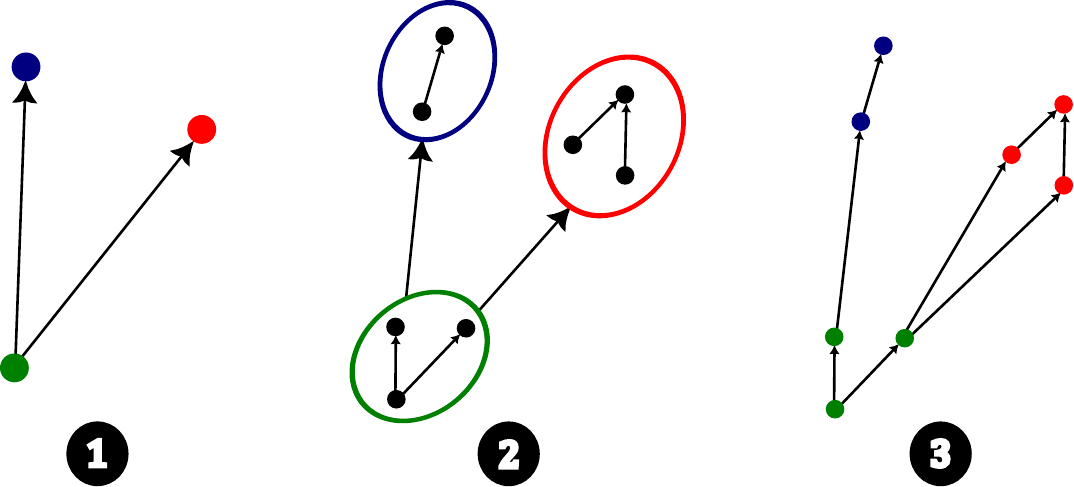}
  \end{minipage}
  \hskip2em
  \begin{minipage}{0.6\linewidth}
    \begin{itemize}[label=$\circ$]
      \item Picture~\Circled{1} shows a 3-element pomset~$P {=} ((E, \leq),
        \lambda)$.
      \item In picture~\Circled{2}, each node~$e \in E$ is now labelled with a
        pomset in the set~$g(\lambda(e))$.
      \item Picture~\Circled{3} shows the final result, where the `inner'
        pomsets have been inlined in~$P$.
    \end{itemize}
  \end{minipage}
  \caption{Substitution of labelled pomset~$P$ with~$g$}
  \label{fig:monad-substitution}
\end{figure}

As discussed in \cref{def:plain-execution}, a plain execution is
simply a pomset (Def. \ref{def:pomset}) labelled with method calls and crashes.
Pomsets are endowed with a natural \emph{substitution operation}\footnote{This
is indeed the `bind' operation of a monad structure on $\Pomset$.}. Consider
a pomset~$P = ((E, \leq), \lambda)$ over a set~$X$ and a map~$g: X \to
\Pomset(Y)$ associating labels in~$X$ with pomsets labelled over a set~$Y$.
Their substitution $P \bind g \in \Pomset(Y)$, depicted in \cref{fig:monad-substitution}, is obtained by replacing each event~$e$ in~$P$
with the pomset~$g(\lambda(e))$.
Formally, the carrier set of the pomset~$P \bind g$ is the disjoint union $F
  \eqdef \coprod_{e \in E} \: \toplain{g (\lambda (e))}$:
each node~$e$ of~$P$ is replaced by the set of nodes of the
pomset~$g(\lambda(e))$.
Elements of this set are of the form~$(e, f)$ with~$f \in \toplain{g (\lambda
    (e))}$, and are ordered lexicographically as follows:
\begin{center}
$
  (e_1, f_1) \leq_{P \cdot g} (e_2, f_2) \iff
  (e_1 \leq_P e_2) \;\vee\; (e_1 = e_2 \land f_1 \leq_{g(\lambda(e_1))} f_2)
$
\end{center}
Finally, the label of an element $(e,f)$ is~$\mu_{\lambda(e)}(f) \in Y$, where
we write $\mu_x: \toplain{g(x)} \to Y$ for the labeling map of the
pomset~$g(x)$.

Because a program is interpreted using a \emph{set} of such pomsets, we also use
the powerset~$\Pow$ which maps a set~$X$ to the set $\{ A \mid A \subseteq X\}$
of its subsets:
$
  \sem{e}^\valbot \;\in\; \Pow\Pomset(\calls(\genlib))
$,
where $\calls(\genlib)$ is the set of \emph{method calls} (of the form
$\methlab{m}{\vec v}{\valbot}{}$) of library~$\genlib$.
We extend the substitution operation to sets of pomsets: given a
set~$\genpomsetset \in \Pow\Pomset(X)$ of pomsets, and a map~$g: X \to
  \Pow\Pomset(Y)$, their substitution is as follows, where the substitution operation used on the right-hand side is the previous one, operating on pomsets, $\Id_S: S \to S$ is the identity map on the set~$S$, and the map~$\lambda'$ has type~$E \to \Pomset(Y)$.
\begin{center}
$
  \genpomsetset \bind g \quad\coloneqq\quad
  \bigcup_{((E, \leq), \lambda) \in \genpomsetset}
  \big\{
  ((E, \leq), \lambda') \bind \Id_{\Pomset(Y)}
  \mid
  \forall e \in E,\: \lambda'(e) \in g(\lambda(e))
  \big\}
$
\end{center}
Informally, a pomset~$P$ belongs to~$\genpomsetset \bind g$ iff it can be
obtained from a pomset $Q = ((E, \leq), \lambda) \in \genpomsetset$ by
replacing each event in~$Q$ by some pomset in~$g(\lambda(e)) \in
\Pow\Pomset(Y)$ -- see \cref{fig:monad-substitution}.

Applying this general operation to plain executions and (semantic)
implementations yields the desired operation because it preserves maximality of
incomplete operations.
\begin{proposition}
  The operation above lifts to an operation on plain executions:
\begin{center}
$
    {-} \cdot \genimpl \;\;:\;\;
    \PExec{\genlib} \;\;\longrightarrow\;\; \Pow(\PExec\coll)
$
\end{center}
\end{proposition}

Another way to formalize how to link an implementation with a program is
syntactically: given a program~$\prog$ and a syntactic
implementation~$\gensynimpl$, we can define~$\prog \bind \gensynimpl$ as the
program where all calls to $\genlib$  methods are replaced with their
source code given by~$\gensynimpl$.
As expected, the two ways of linking with an implementation are
compatible, in the following sense.
\begin{proposition}\label{prop:syntax-semantics}
  Given a program~$\prog$ that uses~$\genlib$ and a syntactic
  implementation~$\gensynimpl$ of~$\genlib$:
\begin{center}
$
    \semprog{\prog} \bind \sem{\gensynimpl}
    =
    \semprog{\prog \bind \gensynimpl}
$
\end{center}
\end{proposition}

\subsubsection{Implementation Correctness}
We now have the notions required to lift the \emph{correctness of an
implementation} to the general weak memory setting.
The main difference with \cref{sec:overview} is that we must consider the two levels
of \emph{plain executions}, which contain the events encountered by the program,
and \emph{executions}, which contain additional information that determine whether the   execution is possible (consistent).

As an execution may contain calls from several libraries, 
given $\genexec \in \Exec\coll$,  $\genlib \in \coll$ and a relation~$r$ on~$\genexec.E$,
we write $\genexec.E_\genlib$ for the $\genexec.E$ events labelled with $\genlib$ method calls, and we write~$r_{\genlib}$ for $r \cap (\genexec.E_\genlib \times \genexec.E_\genlib)$. 
Similarly, we write $\genexec \restrict \genlib$ for $(\genexec.E_\genlib, \genexec.\kpo_\genlib, \genexec.\ksw_\genlib, \genexec.\khb_\genlib)$.

\begin{definition}
  An execution~$\genexec \in \Exec{\coll}$ is \emph{consistent} if:
  \begin{enumerate}
    \item $\genexec.\ksw = \bigcup_{\genlib \in \coll} \genexec.\ksw_\genlib$;
    \item for all~$\genlib \in \coll$, $\genexec \restrict \genlib \in
      \genlib.\libConsistent$; and
    \item for all~$\genlib \in \coll$, $\anonymize{\genlib}{\genexec} \in
      \genlib.\globalConsistent$.
  \end{enumerate}
  Execution $\genexec$ is \emph{hereditarily consistent} if either $\genexec$ is
  empty, or $\genexec$ is consistent and there exists~$\genexec' \prefimm
  \genexec$ such that $\genexec'$ is hereditarily consistent.
\end{definition}
In other words, an execution~$\genexec$ is hereditarily consistent if there
exists a sequence of consistent executions from the empty execution
to~$\genexec$, where each step corresponds to adding one event.

Encapsulation is defined as in \cref{def:encapsulation}, where the
derived happens-before order~$\happensbefore$ is replaced with the
primitive $\khb$ component of the execution.
\begin{definition}\label{def:well-formed-execution}
An execution~$\genexec \in \Exec{\coll}$ is \emph{immediately well-formed} if: 
\begin{enumerate}
  \item $\genexec$ is encapsulated;
  \item $\genexec \restrict \genlib \in \genlib{.}\libWellformed$ for all~$\genlib \in
    \coll$;
  \item $\anonymize{\genlib}{\genexec} \in \genlib{.}\globalWellformed$
    for all $\genlib \in \coll \cup \genlib.\libdeps$.
\end{enumerate}
An  execution~$\genexec$ is \emph{well-formed} if, for all $\genexec'' \prefimm
\genexec' \pref \genexec$, if $\genexec''$ is consistent then $\genexec'$ is
immediately well-formed.
\end{definition}


Intuitively, an implementation~$\genimpl$ of~$\genlib$ over~$\coll$ is \emph{correct}
if
\begin{enumerate*}[label=\arabic*)]
	\item it transports well-formedness from the high-level execution to the low-level one as we expect library~$\genlib$ to be used correctly; and
	\item it transports consistency from the low-level execution to the high-level
one, as we assume the underlying $\coll$ libraries are correctly
implemented.
\end{enumerate*}

\begin{definition}[Correct implementation]\label{def:correct-implementation}
  An implementation~$\genimpl$ of~$\genlib$over~$\coll$ is \emph{correct}, written
  $\libimpl{\coll}{\genimpl}{\genlib}$ if, for all collections~$\coll'$ of
  libraries, all `abstract' plain executions~$\genplainexec \in
  \PExec{\{\genlib\} \cup \coll'}$ and all `concrete' plain executions~$\genplainexec'
  \in \genplainexec \bind \genimpl \subseteq \PExec{\coll \cup \coll'}$:
  \begin{enumerate}
    \item for all well-formed $\genexec$, if $\genexec \refines \genplainexec$, then there exists a well-formed $\genexec'$ such that 
      $\genexec' \refines \genplainexec'$;
    \item for all well-formed and hereditarily consistent $\genexec'$, if $\genexec' \refines \genplainexec'$, then there exists a hereditarily consistent $\genexec$ such that $\genexec \refines \genplainexec$.
  \end{enumerate}
\end{definition}

\noindent
It may not be obvious that the above definition is the right one. In \cref{subsec:framework_program_semantics} below we define the observable semantics of a closed program and prove that this definition is indeed adequate to show that an implementation is correct.

\subsubsection{Semantics of Programs}
\label{subsec:framework_program_semantics}
A program is \emph{well-formed} if all its plain executions can be
justified by well-formed executions:
\begin{center}
$
  \for{\valbot \in \Valbot}\;
  \for{\genplainexec \in \semprog{\prog}^\valbot}\;
  \exists \genexec \refines \genplainexec. \;\;
  \text{ $\genexec$ is well-formed}
$
\end{center}
A program that is \emph{not} well-formed is considered not to have a well-defined behavior, and thus we only consider the semantics of well-formed programs.
Given a well-formed program~$\prog$ that uses libraries~$\coll$, its behavior
is defined as the values justified by a consistent execution:
\begin{center}
$
  \behaviors(\prog)
  \eqdef
  \{
    v \mid \exsts{\genplainexec \in \semprog{\prog}^v} \exsts{\genexec \refines
  \genplainexec} \genexec \text{ is hereditarily consistent}
  \}
$
\end{center}
One justification for the definition of correctness for a semantic
implementation is that it recovers the usual definition of correctness of a
syntactic implementation~\citep{yacovet}.

\begin{theorem}\label{thm:compositional-correctness-is-correct}
  Let~$\prog$ be a well-formed program that uses~$\coll$, and let~$\gensynimpl$ be a syntactic
  implementation of~$\genlib \in \coll$ over~$\coll'$ such
  that~$\sem{\gensynimpl}$ is correct. Then the program~$\prog \bind \gensynimpl$ is
  well-formed and
  $
    \behaviors(\prog \bind \gensynimpl)
    \;\;\subseteq\;\;
    \behaviors(\prog)
 $.
\end{theorem}
\begin{proof}
 We first prove~$\prog \bind \gensynimpl$ is well-formed.
  Let~$\genplainexec' \in \semprog{\prog \bind \gensynimpl}$. According to
  \cref{prop:syntax-semantics} there exists~$\genplainexec \in
  \semprog{\prog}$ such that~$\genplainexec' \in \genplainexec \bind
  \sem{\gensynimpl}$.
  Because $\prog$ is well-formed, there exists a well-formed~$\genexec$ such that $\genexec \refines
  \genplainexec$, and thus we can directly conclude from \cref{def:correct-implementation}.

  Take an arbitrary~$\genval$ such that there is a hereditarily consistent
  $\genexec'$ and $\genexec' \refines \genplainexec' \in \semprog{\prog \cdot \gensynimpl}$.
  As above, we can obtain~$\genplainexec$ such that $\genplainexec' \in
  \genplainexec \bind \sem{\gensynimpl}$, and conclude with the second
  implication of \cref{def:correct-implementation}.
\end{proof}

\subsection{Compositionally Proving Implementation Correctness}
As in \cref{subsec:sc_composition}, we introduce \emph{compositional correctness}, simplifying the global correctness condition in \cref{def:correct-implementation}. This allows us to prove an implementation correct without reasoning about an arbitrary collection~$\coll'$ of other libraries.
Our compositional correctness condition in this general weak memory (consistency) setting is
inspired by the \emph{local soundness} condition in \citep{yacovet}.

\subsubsection{Preliminaries}
We begin with a few definitions that allow us to express
\emph{how} a low-level execution relates to a high-level one by
\emph{matching} a high-level method call with all low-level events that constitute its implementation.
First, we define an operation to transport relations along maps.
\begin{definition}
  Given two sets~$X$ and~$Y$, an irreflexive relation~$r \subseteq X \times X$ and
  a map~$f: X \to Y$, the irreflexive relation~$\existproj{f}{r}$ on~$Y$ is defined
  as follows:
  \begin{center}
  $
    (y_1, y_2) \in \existproj{f}{r} \iff 
    y_1 \neq y_2 \land
    \Exists{x_1 \in \finv{f}(y_1)}
    \Exists{x_2 \in \finv{f}(y_2)}
    (x_1, x_2) \in r
  $
  \end{center}
\end{definition}
Using this operation, we can define \emph{intentionally} when a low-level
plain execution corresponds to `linking' a high-level plain execution with an
implementation as follows.
\begin{definition}
  Given plain executions~$\genplainexec \in \PExec{\genlib}$
  and~$\genplainexec' \in \PExec{\coll}$ and an implementation~$\genimpl$
  of~$\genlib$ over~$\coll$, a map~$f : \genplainexec'.E \to
  \genplainexec.E$ is a \emph{plain matching} if:
  \begin{enumerate}
    \item $f$~is surjective; 
    \item $\genplainexec.\kpo = \existproj{f}{\genplainexec'.\kpo}$; and
    \item $\forall e \in \genplainexec.E. \;\;
      \rest{\genplainexec'}{\finv{f}(e)} \in \genimpl(\lab(e'))$
  \end{enumerate}
  where~$\rest{\genplainexec}{X}$, with $X \subseteq \genplainexec.E$ denotes
  the restriction of~$\genplainexec$ to the set~$X$ of events.
\end{definition}
Intuitively, $f$ denotes that the high-level method call~$e$
in~$\genplainexec$ is implemented using the set of events~$\finv{f}(e)$, and that
the high-level program order is determined by that of the low-level execution.
As captured by following proposition, such a plain matching
\emph{witnesses} the fact that a plain execution results from `applying'
an implementation to a high-level execution.
\begin{proposition}\label{prop:existence-of-surjection}
  Given an implementation $\genimpl$ of~$\genlib$ over~$\coll$ and plain
  executions~$\genplainexec \in \PExec{\genlib}$ and~$\genplainexec' \in
  \PExec{\coll}$, if~$\genplainexec' \in \genplainexec \bind \genimpl$, then
  there exists a matching~$f: \genplainexec' \surjection \genplainexec$.
\end{proposition}

\subsubsection{A compositional criterion}
We state our \emph{compositional correctness} criterion as a lifting
problem as follows.
Given a \emph{plain} matching $f : \toplain{\genexec'} \surjection
\genplainexec$, is it possible to find an execution~$\genexec$ that
refines~$\genplainexec$ ($\genexec \refines \genplainexec$) such that~$f: \genexec' \surjection \genexec$ is a
matching at the level of executions, in the following sense?
\begin{definition}
  Given executions $\genexec$ and~$\genexec'$ and a map~$f$ such that~$f:
  \toplain{\genexec'} \surjection \toplain{\genexec}$, $f$~is a \emph{refined
  matching}, written~$f: \genexec' \surjection \genexec$, if the following hold:
  \begin{enumerate}
    \item $\genexec$ and~$\genexec'$ are consistent;
    \label{item:framework_compositional_consistent}
    \item $\genexec{.}\ksw^+ \subseteq \existproj{f}{\genexec'.\klhb}$; and
    \label{item:framework_compositional_sw}
    \item $\genexec{.}\klhb = (\existproj{f}{\genexec'{.}\klhb \setminus
      (\genexec'.\ksw \cup \genexec'.\kpo)^+} \cup \genexec.\kpo \cup \genexec{.}\ksw)^+$\label{item:framework_compositional_hb}
  \end{enumerate}
\end{definition}
Condition \ref{item:framework_compositional_consistent} captures the intuition that this refined notion of matching lives in an
idealized world, where all executions have good properties.
%
Condition \ref{item:framework_compositional_sw} states that two
high-level events synchronize when there exist two low-level events in
their respective implementations that are related by happens-before.
That is, $\genexec.\ksw$ edges cannot appear out of thin air and must be justified by the implementation.
Condition \ref{item:framework_compositional_hb} states that the high-level happens-before order is determined
by its own $\kpo$ and~$\ksw$ orders as well as the external happens-before
order. The external order is computed by removing the $\genexec'.\kpo$ and $\genexec'.\ksw$ contributions from~$\genexec'.\khb$
and mapping the remaining $\genexec'.\khb$ to~$\genexec$ using~$\existprojop{f}$.
Intuitively, this `external' happens-before is a remnant of other
libraries that are ignored by focussing on library~$\genlib$, and being
compatible with it allows consistency not to depend on these other libraries.

We consider implementations that are \emph{locally well-formed} in that they
preserve well-formedness of `local' executions, \ie executions that only contain
events from either~$\genlib$ for high-level executions or $\coll$ for low-level
ones.
\begin{definition}\label{def:well-formed}
  An implementation $\genimpl$ of~$\genlib$ over~$\coll$ is \emph{locally well-formed} if, for all plain
  executions~$\genplainexec \in \PExec{\genlib}$, all $\genplainexec' \in
  \genplainexec \bind \genimpl \subseteq \PExec{\coll}$ and all $\genexec$, if $\genexec \refines
  \genplainexec$ and $\genexec$ is well-formed, then there exists $\genexec'$ such that $\genexec' \refines \genplainexec'$ and $\genexec'$ is
  well-formed.
\end{definition}

We now state the \emph{compositional correctness} criterion which comprises two parts: a local and a global condition.
The local condition states that the lifting problem mentioned above always
has a solution $\genexec$ such that $\genexec$ corresponds to an immediate prefix~$\prevexec'$ of the low-level execution~$\genexec'$.
This captures an induction on the property that we assume all low-level
executions are hereditarily consistent.
As empty executions are consistent, the base case holds, which means that we
obtain a sequence of refined matchings, and in particular a witness that the
high-level execution is locally consistent.

The global condition then ensures that global consistency predicates of the
dependencies also hold. Consider a dependency~$\genlib' \in \coll$ and
executions~$\genexec' \in \Exec{\coll \cup \{\libtags{\genlib'}\}}$ and~$\genexec \in
\Exec{\{\genlib, \genlib', \libtags{\genlib'}\}}$ with a plain matching between
them.
Restricting $\genexec$ to its $\genlib$ events induces a plain matching, which can be lifted to a refined matching because of the local
condition.
As such, the global condition stipulates that the implementation preserve the global well-formedness and global consistency of~$\genexec$
and~$\genexec'$.

\begin{definition}[Compositional correctness]%
  \label{def:compositional-correctness}
  A well-formed implementation $\genimpl$ of~$\genlib$ over~$\coll$
  is \emph{compositionally correct} if the following two conditions hold.
  
  \begin{enumerate}
    \item Given executions~$\prevexec', \genexec' \!\in\! \Exec{\coll}$ and
      $\prevexec \!\in\! \Exec{\genlib}$ and a plain
      execution~$\genplainexec \in \PExec{\genlib}$,\\
       \textbf{if} $\genexec$ is
      consistent and the following holds\footnote{The right diagram
        should be seen as sitting above the left one, where~$\genexec$ being
      above~$\genplainexec$ means that~$\genexec \refines \genplainexec$.} 
      (\ie \textbf{if} $f$ is a plain matching between $\toplain{\genexec'}$
      and~$\genplainexec$, $\prevexec'$
    is an immediate prefix of~$\genexec'$ and~$\tilde f$ is obtained by
    restricting the domain of~$f$ to~$\prevexec.E$)
    $$
      \begin{tikzcd}
        \toplain{\prevexec'}
        \ar[d, ->>, "\tilde{f}"']
        \ar[r, hook, "\mathit{im}"'{pos=.9}, shorten >=5.5pt]
        &
        \toplain{\genexec'}
        \ar[d, ->>, "f"]
        \\
        \toplain{\prevexec}
        \ar[r, hook, "\mathit{im}"'{pos=.9}, "="{pos=0.9}, shorten >=5.5pt]
        &
        \genplainexec
      \end{tikzcd}
      \qquad\qquad
      \begin{tikzcd}
        \prevexec'
        \ar[d, ->>, "\tilde{f}"']
        \ar[r, hook, "\mathit{im}"'{pos=.9}, shorten >=5.5pt]
        &
        \genexec'
        \\
        \prevexec
      \end{tikzcd}
   $$
    \textbf{then} there exists an execution~$\genexec$ such that $\genexec \refines \genplainexec$ and the following
    holds\footnote{This amounts to completing the diagram sitting on top to
    obtain a cube.}:
    $$
      \begin{tikzcd}
        \prevexec'
        \ar[d, ->>, "\tilde{f}"']
        \ar[r, hook, "\mathit{im}"'{pos=.9}, shorten >=5.5pt]
        &
        \genexec'
        \ar[d, ->>, "f"]
        \\
        \prevexec
        \ar[r, hook, "\mathit{im}"'{pos=.9}, "="{pos=0.9}, shorten >=5.5pt]
        &
        \genexec
      \end{tikzcd}
    $$
  \item \label{item:framework_compositional_correctness_wf}
  For all~$\genlib' \in \coll$, given executions~$\genexec' \!\in\!
    \Exec{\coll \cup \{\libtags{\genlib'}\}}$ and~$\genexec \!\in\! \Exec{\{\genlib, \genlib',
    \libtags{\genlib'}\}}$, consider $f: \toplain{\genexec'} \surjection
    \toplain{\genexec}$ such that for any event~$e$
    not from~$\genlib$, $\finv{f}(e)$ is a singleton.
    Let~$\genexec'_{\genlib'} \eqdef \finv{f}(\genexec \restrict \genlib')$
    and assume~$f$ restricts to~$f_{\genlib'}: \genexec'_{\genlib'} \surjection
    \genexec \restrict \genlib'$.
    Then the following must hold:
    \begin{itemize}
      \item if $\anonymize{\genlib'}{\genexec} \in \genlib'.\globalWellformed$,
        then~$\anonymize{\genlib'}{\genexec'} \in \genlib'.\globalWellformed$; and
      \item if $\anonymize{\genlib'}{\genexec'} \in \genlib'.\globalConsistent$,
        then~$\anonymize{\genlib'}{\genexec} \in \genlib'.\globalConsistent$.
    \end{itemize}
  \end{enumerate}
\end{definition}

\begin{theorem}
  Compositional correctness implies correctness.
\end{theorem}

\section{Case Study: Durable Linearizability with \Flit and \Mirror}
\label{sec:flit_and_mirror}
We consider a family of libraries that provide a simple interface with persistent memory accesses (reads and writes), allowing one to convert any linearisable implementation to a durably linearisable one by replacing regular (volatile) accesses with persistent ones supplied by the library. 
Specifically, we consider two such libraries \Flit~\cite{Flit} and \Mirror~\cite{Mirror}; 
we specify them both in our framework, prove their implementations sound against their respective specifications, and further prove their general result for converting data
structures.

\subsection{The \Flit Library}

\begin{figure}
  \begin{minipage}[t]{0.49\linewidth}
    \footnotesize
    \begin{tabular}{l}
      \textbf{method}\ $\flitWrite{\pi}(\ell, v):$
      \\\quad  \textbf{if} $\pi = \flitPersistent$ \textbf{then}
      \\\qquad        $\textsf{fetch-and-add}(\textit{flit-counter}(\ell), 1)$;
      \\\qquad        $\store{\ell}{v}$;
      \\\qquad        $\foC \ell$;
      \\\qquad        $\textsf{fetch-and-add}(\textit{flit-counter}(\ell), -1)$;
      \\\quad  \textbf{else}
      \\\qquad        \sfenceC;
      \\\qquad        $\store{\ell}{v}$;
    \end{tabular}
  \end{minipage}\begin{minipage}[t]{0.49\linewidth}
    \footnotesize
    \begin{tabular}{l}
      \textbf{method} $\flitRead{\pi}(\ell):$
      \\\quad    \textbf{local} $v = \load{\ell}$;
      \\\quad    \textbf{if} $\pi = \flitPersistent \land \textit{flit-counter}(\ell) > 0$ \textbf{then}
      \\\qquad        $\foC \ell$;
      \\\quad    \textbf{return} $v$;
      \\
      \\\textbf{method} $\finishOp:$
      \\\quad        \sfenceC;
    \end{tabular}
  \end{minipage}\vspace{-10pt}
  \caption{$\Flit$ library implementation in \pxes 
  }
  \label{fig:flit}
\end{figure}

\newcommand{\flitP}{\flitPersistent}
\newcommand{\flitV}{\flitVolatile}

\Flit \citep{Flit} is a persistent library that provides a simple interface very close
to \pxes, but with stronger persistency guarantees, which make it easier to
implement durable data structures.
Specifically, a \Flit object $\ell$ can be accessed via write and read methods,
$\flitWrite{\pi}(\ell, v)$ and $\flitRead{\pi}(\ell)$, as well as standard
read-modify-write methods.
Each write (\resp read) operation has two variants, denoted by the \emph{type}
$\pi \in \{\flitPersistent, \flitVolatile \}$. This type specifies if the write
(\resp read) is \emph{persistent} ($\pi = \flitPersistent$) in that its effects
must be persisted, or \emph{volatile} ($\pi =\flitVolatile$) in that its
persistency has been optimised and offers weaker guarantees.
The default access type is persistent ($\flitPersistent$), and the volatile accesses may be used as optimizations when weaker guarantees suffice. Wei \etal \cite{Flit} introduce a notion of \emph{dependency} between different operations as follows. If a (persistent
or volatile) write $w$ depends on a persistent write $w'$, then $w'$ persists
before $w$. If a persistent read $r$ reads from a persistent write $w$, then
$r$ depends on $w$ and thus $w$ must be persisted upon reading if it has not
already persisted.
%
%
Though simple, \Flit provides a strong guarantee as captured by a general
result for correctly converting volatile data structures to persistent ones: if
one replaces every memory access in the implementation of a \emph{linearizable}
data-structure with the corresponding persistent \Flit{} access, then the
resulting data structure is \emph{durably linearizable}.

Compared to the original \Flit development, our soundness proof is more formal
and detailed: it is established against a formal specification (rather than an
English description) and with respect to the formal \pxes model.

\paragraph{\Flit Interface}
The \Flit interface uses the~$\Ptag$ from~$\pxes$ and contains a single
constructor, \textsf{new}, allocating a new \Flit location, as well as three
other methods below, the last two of which are durable:
\begin{itemize}
  \item $\flitRead{\pi}(\ell)$ with $\pi \!\in\! \{\flitPersistent, \flitVolatile\}$, for a $\pi$-read from $\ell$;
  \item $\flitWrite{\pi}(\ell, v)$ with $\pi \!\in\! \{\flitPersistent, \flitVolatile\}$, denoting a $\pi$-write of value $v \in \Val$ to $\ell$; and
   \item $\finishOp$, which waits for previously executed operations to persist.
\end{itemize}
We write $R$ and $W$ respectively for the read and write events, and add the superscript $\pi$ (\eg $R^{\flitPersistent}$) to denote such events with the given persistency mode.

\paragraph{\Flit Specification}
We develop a formal specification of \Flit in our framework, based on its
original informal description. The correctness of \Flit executions is described
via a \emph{dependency} relation that contains the program order and the total
execution (linearization) order restricted to persistent write-read operations
on the same location. Note that this dependency notion is stronger than the
customary definitions that use a $\kcom$ relation (as in the \pxes specification)
instead of $\klin$, because a persistent read may not read directly from a
persistent write $w$, but rather from another later ($\klin$-after $w$) write.

\begin{definition}[\Flit execution Correctness]
  A \Flit execution $\genexec$ is correct if there exists a `reads-from'
  relation~$\kcom$ and a total order $\klin \supseteq \genexec.\klhb$ on
  $\genexec.E$ and an order~$\knvo$ such that:
  \begin{enumerate}
    \item Each read event reads from the most recent previous write to the same
          location:\\ $\kcom = \bigcup_{\ell\in\Loc} ([W_\ell] ; \klin ;
            [R_\ell]) \setminus (\klin ; [W_\ell] ; \klin)$
    \item Reads return the value written by the write they read from:\\ $(w,r) \in
            \kcom \Rightarrow \exsts{\ell, \pi, \pi', v} \lab(r) =
            \flitRead{\pi'}(\ell):v \land \lab(w) = \flitWrite{\pi}(\ell, v):-$
    \item Persistent writes persist before every other later dependent write:\\
          $[W^\flitP] ; (\kpo \cup \bigcup_{\ell\in\Loc} [W^\flitP_\ell] ; \klin ;
            [R^\flitP_\ell])^+ ; [W] \subseteq \knvo$
    \item Persistent writes before a $\finishOp$ persist:\\ $\dom([W^\flitP] ; (\kpo \cup
            \bigcup_{\ell\in\Loc} [W^\flitP_\ell] ; \klin ; [R^\flitP_\ell])^+ ;
            [\finishOp]) \subseteq \makeset{\Ptag}$
    \item And $\knvo$ is a persist order: $\dom(\knvo; \makeset{Ptag}) \subseteq
      \makeset{Ptag}$.
  \end{enumerate}
\end{definition}

\paragraph{\pxes implementation of \Flit}
The implementation of \Flit methods is given in \cref{fig:flit}. Whereas a
naive implementation of this interface would have to issue a flush instruction
both after persistent writes and in persistent reads, the implementation shown
associates each location with a counter to avoid performing superfluous flushes
when reading from a location whose value has already persisted. Specifically, a
persistent write on $\ell$ increments its counter before writing to and
flushing it, and decrements the counter afterwards. As such, persistent reads
only need to issue a flush if the counter is positive (\ie if there is a
concurrent write that has not executed its flush yet).


\begin{restatable}{theorem}{flitCorrectTheorem}\label{thm:flit-correct}
  The implementation of \Flit in \cref{fig:flit} is correct.
\end{restatable}

\paragraph{\Flit and Durable Linearizability}
Given a data structure implementation $I$, let $p(I)$ denote the
implementation obtained from $I$ by
\begin{enumerate*}[label=\arabic*)]
  \item replacing reads/writes in the implementation with
        their corresponding persistent \Flit instructions, and
  \item adding a call to $\finishOp$ right before the end of each method.
\end{enumerate*}
We then show that given an implementation $I$, if $I$ is linearizable, then
$p(I)$ is \emph{durably linearizable}\footnote{The definition here is the same
as in \cref{sec:overview}, as $\khb$-linearizations of the
execution still yield sequential executions.}.
%
We assume that all method implementations are single-threaded, \ie all plain
executions~$I(m(\vec v))$ are totally ordered.

\begin{restatable}{theorem}{flitLinTheorem}\label{thm:flit-durlin}
  If $\libimplC{\pxes}{I}{\Lin(S)}$, then $\libimplC{\Flit}{p(I)}{\DurLin(S)}$.
\end{restatable}

\subsection{The \Mirror Library}
The Mirror \citep{Mirror} persistent library has similar goals to \Flit. 
The main difference between the two is that \Mirror
operations do not offer two variants, and their operations are implemented
differently from those of \Flit. Specifically, in \Mirror each location has two
copies: one in persistent memory to ensure durability, and one in volatile
memory for fast access. As such, read operations are implemented as simple loads
from volatile memory, while writes have a more involved implementation than
those of \Flit.

We present the \Mirror specification and implementation in the technical appendix where we also prove that its implementation is correct against its specification. 
As with \Flit, we further prove that \Mirror can be used to convert linearizable data structures to durably linearizable ones, as described above.

\section{Case Study: Persistent Transactional Library}
\label{sec:transactions}

We revisit the $\Ltrans$ transactional library, develop its formal specification
and verify its implementation (\cref{fig:Ltrans-implementation}) against it. 
%
%
Recall the simple $\Ltrans$ implementation in \cref{fig:Ltrans-implementation}
and that we do not allow for nested transactions. 
%
The implementation uses an \emph{undo-log} which records the former values of
persistent registers (locations) modified in a transaction. If, after a crash,
the recovery mechanism detects a partially persisted transaction (\ie the last
entry in the undo log is not \lstinline|COMMITTED|), then it can use the undo-log
to restore registers to their former values. 
The implementation uses a durably linearizable queue
library\footnote{For example, take any linearizable queue implementation and use the \Flit library as described in \cref{sec:flit_and_mirror}.}~$\makeMM{Q}$, and assumes that it is \emph{externally synchronized}: the user is
responsible for ensuring no two transactions are executed in parallel. We
formalize this using a global well-formedness condition.

Later in \cref{subsec:transactions_lock} we develop a wrapper library~$\LStrans$
for $\Ltrans$ that additionally provides synchronization using locks and prove
that our implementation of this library is correct. 
To do this, we need to make small modifications to the structure of the specification: the specification in \cref{sec:overview} requires that any
`transaction-aware operation' (\ie those tagged with~$\Ttag$) be enclosed in calls to $\PTBegin$ and~$\PTEnd$.
Since~$\LStrans$ wraps the calls to~$\PTBegin$ and~$\PTEnd$, the well-formedness
condition needs to be generalized to allow operations tagged with~$\Ttag$ to
appear between calls to operations that behave like~$\PTBegin$ and~$\PTEnd$.
To that end, we add two new tags~$\Btag$ and~$\Etag$ to denote such operations, respectively.

\subsection{Specification}
\label{subsec:transactions_spec}

The $\Ltrans$ library provides four \emph{tags}: 
\begin{enumerate*}[label=\arabic*)]
  \item $\Ttag$ for transaction-aware `client' operations;
  \item $\perTag$ for operations that have persisted using \textsf{tr}ansactions; and
  \item $\Btag, \Etag$ for operations that begin and end transactions,
    respectively.
\end{enumerate*}
We write $\PTReadSet, \PTWriteSet, \BeginSet, \EndSet, \RecoverSet$ respectively
for the sets of events labeled with read, write, begin, end and recovery
methods.
As before, we write \eg $\Ttagset$ for the set of events tagged with~$\Ttag$.
Note that while $\BeginSet$ denotes the set of the begin events in library
$\Ltrans$, the $\Btagset$ denotes the set of all events that are tagged with
$\Btag$, which includes $\BeginSet$ (of library $\Ltrans$) as well as events of
\emph{other} (non-$\Ltrans$) libraries that may be tagged with $\Btag$;
similarly for $\EndSet$ and $\Etagset$. As such, our local specifications below
(\ie local well-formedness and consistency) are defined in terms of $\BeginSet$
and $\EndSet$, whereas our global specifications are defined in terms of
$\Btagset$ and $\Etagset$. As before, for brevity we write \eg
$\diagonal{\Ttag}$ as a shorthand for the relation $\diagonal{\makeset\Ttag}$.
We next define the `same-transaction' relation $\ksamet$:
\[
  \ksamet \eqdef \diagonal{\Btagset \cup \Etagset
  \cup \Ttagset}; (\kpo\cup\kpo^{-1}); \diagonal{\Btagset \cup \Etagset \cup \Ttagset}
  \;\;\setminus\;\; ((\kpo; \diagonal\Etag; \kpo) \cup (\kpo; \diagonal\Btag; \kpo))
\]
%

\noindent An execution is locally well-formed iff the following hold: 
\begin{enumerate}
  \item\label{PT:open-close} A transaction must be opened before it is
    closed: $\EndSet \subseteq \rng([\BeginSet] ; \kpo)$
  \item\label{PT:sections-match} Transactions are not nested and are
    matching: $[\EndSet]; \kpo; [\EndSet] \subseteq [\EndSet]; \kpo; [\BeginSet];
    \kpo; [\EndSet]$ and $[\BeginSet]; \kpo; [\BeginSet] \subseteq [\BeginSet];
    \kpo; [\EndSet]; \kpo; [\BeginSet]$
  \item\label{PT:sections-sync} Transactions must be externally
    synchronized: $\EndSet \times \BeginSet \subseteq \khb \cup \khb^{-1}$
  \item\label{PT:recovery} The recovery routine must be called after a crash:
    $\Crash; \khb; \Btagset \subseteq \Crash; \khb; [\RecoverSet]; \khb;
    \Btagset$
  \item Events are correctly tagged: $\mathcal{W} \cup \mathcal{R} \subseteq
    \makeset{\Ttag}$
\end{enumerate}
An execution is globally well-formed if client operations ($\Ttag$-tagged) are
inside transactions:
\begin{enumerate}[resume]
  \item\label{PT:PT-range} $\Ttagset \subseteq \rng(\diagonal\Btag; \kpo)$ 
  \item\label{PT:PT-po} $\diagonal\Etag ; \kpo ; \diagonal\Ttag \subseteq \diagonal\Etag; \kpo; \diagonal\Btag; \kpo; \diagonal\Ttag$ 
\end{enumerate}
An execution is locally-consistent if there exists a `reads-from'
relation~$\kcom$ such that: 
\begin{enumerate}[resume]
  \item\label{PT:rf-read-write} $\kcom$ relates writes to reads, $\krf \subseteq
    \PTWriteSet \times \PTReadSet$, such that each read is related to exactly one write (\ie $\finv{\kcom}$ is total and functional).
  \item\label{PT:recent-reads} Reads access the most recent write: $\inv\kcom;
    \khb \subseteq \khb$
  \item\label{PT:persist-external-reads} 
   External reads (reading from a different transaction) read from persisted writes: 
  $\dom(\kcom \setminus \ksamet) \subseteq \makeset{\perTag}$
\end{enumerate}
An execution is globally-consistent if there exists an order~$\knvo$ over
$\makeset\Ttag$ such that:
\begin{enumerate}[resume]
  \item\label{PT:hb-nvo} Transactions are $\knvo$-ordered: $\diagonal\Etag ; \khb ;
    \diagonal\Btag \subseteq \knvo$ 
  \item $\knvo$ is the persistance order: $\dom(\knvo ; \diagonal{\perTag})
    \subseteq \makeset{\perTag}$;
  \item\label{PT:P-st} Either all or none of the events in a transaction persist (atomicity): 
    $\diagonal{\perTag}; \ksamet; \diagonal\Ttag \subseteq \diagonal{\perTag}$ 
  \item\label{PT:P-complete} 
  All events of a completed transaction (ones with an associated end event)
  persist: 	  $\Etagset^c \subseteq \makeset{\perTag}$,  where $\Etagset^c$
  denotes the set of method calls tagged with~$\Etag$ which have completed.
\end{enumerate}


\begin{theorem}
The $\Ltrans$ implementation in \cref{fig:Ltrans-implementation} over~\pxes is correct. 
%
\end{theorem}

\subsection{Vertical Library Composition: Adding Internal Synchronization}
\label{subsec:transactions_lock}
We next demonstrate how our framework can be used for \emph{vertical library
composition}, where an implementation of one library comprises calls to other
libraries with non-trivial global specifications.
To this end, we develop $\LStrans$, a wrapper library around~$\Ltrans$ that is
meant to be simpler to use by providing synchronization internally: rather than
the user ensuring synchronization for $\Ltrans$, one can use $\LStrans$ to
prevent two transactions from executing in parallel. 
More formally, the well-formedness condition \eqref{PT:sections-sync}
of~$\Ltrans$ becomes a correctness guarantee of~$\LStrans$.
We consider a simple implementation of~$\LStrans$ that uses a global lock
acquired at the beginning of each transaction and released at the end as shown
below.
%
\begin{lstlisting}[language=Imp]
globals lock := L.new()         method LPTBegin() := L.acq(lock);PTBegin()   
                                method LPTEnd() := PTEnd();L.rel(lock)
\end{lstlisting}

\begin{theorem}
  The implementation of $\LStrans$ above is correct.
\end{theorem}
Using compositional correctness, the main proof obligation is condition \ref{item:framework_compositional_correctness_wf} of
\cref{def:compositional-correctness} stipulating that the implementation be
well-formed, ensuring that~$\Ltrans$ is used correctly by the $\LStrans$
implementation.
This is straightforward as we can assume there exists an immediate prefix that is
consistent (\cref{def:well-formed-execution}). The existence of the
$\khb$-ordering of calls to~$\PTBegin$ and $\PTEnd$ follows from the
consistency of the global lock used by the implementation.

\subsection{Horizontal Library Composition}
We next demonstrate how our framework can be used for \emph{horizontal library
composition}, where a \emph{client} program comprises calls to multiple
libraries. To this end, we develop a simple library, $\Counterlib$, providing a
persistent counter to be used in \emph{sequential} (single-threaded) settings.
As such, if a client intends to use $\Counterlib$ in concurrent settings, they
must call its methods only within critical sections. 
%
The $\Counterlib$ provides three operations to create (\lstinline|NewCounter|),
increment (\lstinline|CounterInc|) and read a counter (\lstinline|CounterRead|).
The specification and implementation of $\Counterlib$ are given in the technical
appendix.

As $\Counterlib$ uses the tags of~$\Ltrans$, we define $\Counterlib{.}\libdeps
\eqdef \{ \Ltrans \}$. The all the operations are tagged with~$\Ttag$.
As such, $\Counterlib$ inherits the global well-formedness condition
of~$\Ltrans$, meaning that $\Counterlib$ operations must be used within
transactions (\ie $\khb$-between operations respectively tagged with~$\Btag$
and~$\Etag$).
Putting it all together, the following client code snippet uses $\Counterlib$ in
a correct way, even though~$\Counterlib$ has no knowledge of the existence
of~$\LStrans$.
\begin{lstlisting}[language=Imp]
    c = NewCounter();  LPTBegin();  CounterInc(c);  CounterInc(c);  LPTEnd();
\end{lstlisting}
Specifically, the above is an instance of horizontal library composition (as the
client comprises calls to both $\LStrans$ and $\Counterlib$), facilitated in our
framework through global specifications.

\section{Conclusions, Related and Future Work}
\label{sec:conclusions}
We presented a framework for specifying and verifying
persistent libraries, and demonstrated its utility and generality by encoding
existing correctness notions within it and proving the
correctness of the \Flit and \Mirror libraries, as well as a persistent
transactional library.

\paragraph{Related Work}
The most closely related body of work to ours is \cite{yacovet}.
However, while their framework can be used for specifying only the consistency
guarantees of a library, ours can be used to specify both consistency and
persistency guarantees. Existing literature includes several works on formal
persistency models, both for hardware
\cite{epochstrand,Px86,parm-oopsla,parmv8,persist-buffering,TamingPx86,ptso,NTSx86}
and software \cite{atlas,arp,sfr}, as well as correctness conditions for
persistent libraries such as durable linearizability \cite{Izraelevitz}. As we
showed in \cref{sec:framework}, such models can be
specified in our framework.


Additionally, there are several works on implementing and verifying algorithms
that operate on NVM. \citep{pds-pmsq} and \citep{pds-set} respectively
developed persistent queue and set implementations in \pxes.
\citep{persistent-queue-proof} provided a formal correctness proof of the
implementation in \cite{pds-set}. All three of
\cite{persistent-queue-proof,pds-set,pds-pmsq} assume that the underlying
concurrency model is SC \cite{lamport-sc}, rather than that of \pxes (namely
TSO). As we demonstrated in \cref{sec:flit_and_mirror}--\cref{sec:transactions} we can
use our framework to verify persistent implementations \emph{modularly} while
remaining faithful to the underlying concurrency model. \citep{pog,pierogi} have
developed persistent program logics for verifying programs under \pxes.
\citep{persevere} recently formalized the consistency and persistency semantics
of the Linux ext4 file system, and developed a model-checking algorithm and tool
for verifying the consistency and persistency behaviors of ext4 applications
such as text editors.

\paragraph{Future Work}
We believe our framework will pave the way for further work on verifying
persistent libraries, whether manually (as done here),
possibly with the assistance of an interactive theorem prover and/or program
logics such as those of \citep{compass,pog,pierogi}, or automatically via model
checking.
The work of \citep{compass} uses the framework of \citep{yacovet}
to specify data structures in a program logic, and it would be
natural to extend it to our framework for persistency.
Existing work in the latter research direction, \eg \cite{demsky,persevere}, has
so far only considered low-level properties, such as the absence of races or the
preservation of user-supplied invariants. It has not yet considered higher-level
functional correctness properties, such as durable linearizability and its
variants. We believe our framework will be helpful in that regard.
In a more theoretical direction, it would be interesting to understand how our
compositional correctness theorem fits in general settings for abstract logical relations such
as \citep{logical-relations}.

\bibliography{../references}

\clearpage
\appendix

\section{Compositional soundness}

We prove \cref{thm:compositional-correctness-is-correct}, the adequacy of the
compositional correctness criterion.
We first prove the following lemma:
\begin{lemma}\label{lem:restriction}
  Consider an implementation~$\genimpl$ of~$\genlib$ on top of~$\coll$ and let
  $\genexec'$ be a consistent and closed execution of~$\Lambda', \Lambda$.
  and~$\genplainexec'$ be a plain execution of~$\Lambda', \genlib$, such that
  $\toplain{\genexec'} \in \genplainexec \bind I$.
  Let $f: \toplain{\genexec'} \to \genplainexec $ be the map given by
  Prop~\ref{prop:existence-of-surjection}. Let $G_L$ be the restriction
  of~$\genplainexec$ to events in~$\genlib$ and $\genexec'_L \coloneqq
  \rest{\genexec'}{\finv{f}(\genplainexec_L.E)}$ be the corresponding
  restriction of~$\genexec'$.
  Let~$f_L : \toplain{\genexec'_L} \surjection \genplainexec_L$ be the
  corresponding restriction of~$f$. 

  If there exists a lifting $f_L : \genexec'_L \surjection \genexec_L$ with
  $\genexec_L \refines \genplainexec_L$, then the plain execution $G'$ can be
  lifted to a closed and locally consistent execution~$\genexec' \refines G'$
  such that $\genexec'{.}\khb \subseteq \existproj{f}{\genexec.\khb}$. In
  particular, $(\kpo' \cup \ksw')^+$ is acyclic.
\end{lemma}
\begin{proof}
  First, let us define an execution~$\genexec$ refining~$\genplainexec$
  using~$\genexec'_L$, and~$\genexec$.
  Let $\ksw$ be $(\existproj{f}{\genexec'{.}\ksw}_\Lambda') \cup
  \genexec_L{.}\ksw$ (we use implicitly the inclusion $G_L \to G$).
  Define
  \[
    \genexec \;\coloneqq\; \tuple{G, \ksw, (\kpo \cup \ksw)^+}
  \]
  To prove that~$\genexec'_L$ is well-formed according to the specification of
  the library~$\genlib'$, it suffices to prove that~$\genexec'{.}\klhb \restrict
  \genlib'$ is acyclic for $\genlib' \in \Lambda' \cup \{ \genlib \}$.
  As we assume that~$\genexec'$ is consistent, it suffices to prove
  that~$\genexec'{.}\klhb = (\kpo' \cup \ksw')^+$ is acyclic.
  We assume by contradiction that there exists a cycle~$x_1, \ldots, x_n = x_1$
  in this relation; the contradition will be the existence of a cycle
  in~$\genexec_L{.}\klhb$.

  Because~$f$ is surjective, for all~$i$, there are three possible cases
  (in this proof, primed relations are components of~$\genexec'$ and unprimed
  are components of~$\genexec$):
  \begin{enumerate}
    \item $x_i \xrightarrow{\kpo} x_{i+1}$, and then there exist~$y'_i$
      and~$y_{i+1}$ in~$G'$ such that $y'_i \xrightarrow{\kpo'} y_{i+1}$
      and~$f(y'_i) \neq f(y_{i+1})$;
    \item $x_i \xrightarrow{\existproj{f}{\ksw'} \cap \genlib^c} x_{i+1}$, then
      $x_i, x_{i+1} \notin \genlib$, and there exist $y'_i$
      and~$y_{i+1}$ in~$G'$ such that $y'_i \xrightarrow{\ksw'} y_{i+1}$;
    \item $x_i \xrightarrow{\ksw_L} x_{i+1}$, then $x_i, x_{i+1} \in \genlib$,
      and there exist~$y'_i$ and~$y_{i+1}$ in~$G'_L$ such that $f(y'_i)
      \xrightarrow{\ksw_L} f(y_{i+1})$.
  \end{enumerate}
  Moreover, the~$y$'s can be chosen such that~$f(y'_i) = f(y_i) = x_i$.
  If none of the~$x_i$ are in~$\genlib$, it is easy to construct a cycle in~$G$,
  and so we can assume without loss of generality that~$x_1 \in \genlib$.

  Consider now the subsequence~$x_{\sigma(j)}$ of the elements of~$(x_i)$ which
  are in~$\genlib$.
  Then the $z_j := y_{\sigma(j)}$ and the $z_j := y'_{\sigma(j)}$ are in~$G_L$,
  and, for each $j$, there are three cases:
  \begin{enumerate}
    \item $z'_j \xrightarrow{\kpo'} z_{j+1}$,
    \item $z'_j \xrightarrow{\finv{f}(\ksw_L)} z_{j+1}$,
    \item $z'_j \xrightarrow{\klhb' \setminus (\ksw' \cup \kpo')} z_{j+1}$
  \end{enumerate}
  Thus, the sequence $f(z_j)$ is a cycle in~$\genexec'_L$ for the relation
  $\existproj{f}{\genexec'{.}\klhb \setminus (\ksw' \cup \kpo')} \cup \kpo \cup
  \ksw$, which contradicts the fact that $\genexec_L$ is well-formed.
\end{proof}

We can now prove \cref{thm:compositional-correctness-is-correct}:
\begin{proof}
  We consider a well-formed implementation~$\genimpl$ of~$\genlib$ on top
  of~$\coll$, as well as a collection~$\coll'$ of libraries and plain executions
  $\genplainexec \in \PExec{\{\genlib\} \cup \coll}$ and~$\genplainexec' \in
  \genplainexec \bind \genimpl \subseteq \PExec{\coll \cup \coll'}$.
  The first part of \cref{def:correct-implementation} follows immediately from
  the fact that~$\genimpl$ is well-formed.
  Let us then prove the second part, and consider~$\genexec' \refines
  \genplainexec'$ well-formed, consistent and closed.
  We need to find~$\genexec \refines \genplainexec$ which is consistent and
  closed.
  Since $\genexec'$ is hereditarily consistent, there exists a sequence
  \[
    \emptyset = \genexec'_0 \prefimm \genexec'_1 \prefimm \cdots \prefimm
    \genexec'_n = \genexec'
  \]
  of consistent closed executions of~$\coll \cup \coll'$.
  As above, by considering the images of these subexecutions under~$f$, we
  obtain
  \[
    \emptyset = \genplainexec_0 \prefimm^= \genplainexec_1 \prefimm^= \cdots \prefimm^=
    \genplainexec_n = \genplainexec
  \]
  with~$f_i: \toplain{\genexec_i} \surjection \genplainexec_i$.
  Further, by taking the restrictions of~$\genplainexec_i$ to the
  library~$\genexec$, we obtain a sequence of~$f_i^\genlib:
  \toplain{\genexec'^\genlib_i} \surjection \genplainexec_i^\genlib$.
  The first plain matching~$f_0^\genlib$ lifts trivially to a refined matching,
  and applying the assumption that~$\genimpl$ is sound~$n$ times, we lift all
  the~$f_i^\genlib$ to refined mathchings, and we get a sequence of consistent
  executions:
  \[
    \emptyset = \genexec_0^\genlib \prefimm^= \genexec_1^\genlib \prefimm^= \cdots \prefimm^=
    \genexec_n^\genlib = \genexec^\genlib
  \]
  Applying the lemma above, we get a sequence of consistent and closed
  executions~$\genexec_i \refines \genplainexec_i$, which are related
  by~$\prefimm^=$ because of the restriction to~$\ksw$ in the definition of
  refined matching and the restriction on~$\kpo$ in the definition of plain
  matching. This concludes the proof.
\end{proof}

\section{Correctness of the \Flit Implementation}
\label{app:flit_proof}

\flitCorrectTheorem*
\begin{proof} 
  We use our modularity theorem. Let $\genexec' = \tuple{E', \kpo', \ksw',
  \klhb'}$ be a consistent \pxes execution, and let~$\kcom'$ and~$\ktso$ be
  witnesses. $\genplainexec = \tuple{E, \kpo}$ be a plain \Flit execution and
  let $f : \toplain{\genexec'} \surjection \genplainexec$ be a plain matching.
  We then define $\klin \eqdef \existproj{f}{\ktso'_M}$, where $\tso_M$ denotes
  the restriction of $\ktso$ to events on location~$\ell$.
  The order $\klin$ is total, because a read $\kpo$-after a write to the same
  location is also ordered by~$\ktso$ thanks to the fetch-and-add operation
  after the write.
  We now prove that the relation~$\klin$ satisfies the four properties in the
  specification of \Flit{}.

  \emph{Proof of}~(1) \emph{and}~(2). The $\klin$-maximal write to the location
  is the $\ktso'$ maximal write to that location before a crash, or the
  $\knvo'$-maximal such write before a crash.

  \emph{Proof of}~(3).  
  First, notice that $\kpo \subseteq \knvo$, because of the flush and the
  barrier.
  Moreover, $[W^\flitPersistent_\ell]; \project{f}{\ktso'_M};
  [R^\flitPersistent_\ell]; \kpo'; [W] \subseteq \knvo$: write~$w, r, w'$ for
  the two writes and the read which are mentioned in this expression.
  There are several cases.
  First, notice that if the two writes are executed on
  on the same thread, then this follows from the first point.
  There are now two cases remaining, depending on whether the read operation
  goes though its fast path, or whether it executes the optimized flush.
  \begin{enumerate}
    \item \emph{Fast path.} If the read operation reads~$0$ in the
      flit-counter, then the write it read must have been $\kmo'$-after or
      equal to the decreasing FAA operation executed by the write operation
      which was read.
      In particular, this means that any write which is $\kpo'$-after that read
      is $\ktso'$-after, and therefore $\knvo'$-after, the decreasing FAA, which
      is $\knvo'$-after the first write.
    \item \emph{Slow path.} Any write which is $\ktso'$-after the read operation
      is $\knvo$-after its optimized flush, which is $\ktso'$-after the write
      operation.
  \end{enumerate}
  Then, the final write executes either a fence or an RMW before its
  linearization point, so that $(r, w') \in \knvo$.

  \emph{Proof of}~(4). Follows from the properties of flushes in \pxes.
\end{proof}

In this proof, it is useful to change the causal structures of executions to
make incomplete events maximal in order to create a crashless execution from a
chain. Given a plain execution $\tuple{E, \kpo}$ and a set of events $E'
\subseteq E$, we define $\dettached{E'}{\tuple{E, \kpo}} \eqdef \tuple{E, \kpo
\setminus (E' \times (E \setminus E'))}$, \ie $E'$ are made $\kpo$-maximal in
$\dettached{E'}{\tuple{E, \kpo}}$.

\flitLinTheorem*
\begin{proof}
  Consider a plain execution~$\genplainexec$ of~$\DurLin(S)$, and a
  corresponding consistent program execution~$\genexec'$ such that
  $\toplain{\genexec'} \in \genplainexec \bind p(I)$.
  Let~$\kcom'$ be a witness that~$\genexec'$ is a consistent \pxes execution.
  There exists a plain matching~$f: \toplain{\genexec'} \surjection
  \tilde{\genplainexec}$.

  To use the fact that~$I$ implements~$\Lin(S)$, we construct an execution
  (without crashes) of~$I$ and a corresponding plain execution of~$\Lin(S)$.
  Consider $\bar \genexec'$ defined as follows.
  \begin{enumerate}
    \item First, detach all the~$\finv{f}(m(\vec v, \bot))$ in~$\genexec'$, giving the
          plain execution~$\genexec'^d$;
    \item Second, define $\genexec'^p \eqdef \genexec'^d \cap
      \downarrow_\kpo \makeset{\Ptag}$, where $\downarrow_\kpo \makeset{\Ptag}$ is the set of
      events of~$\genexec'^d$ which are $\kpo$-before a persisted \pxes event;
    \item Remove all $\finishOp$ events, and relabel all p-writes and p-reads with the
          corresponding \pxes labels;
    \item Finally, we get $\bar \genexec'$ by removing all crash events.
  \end{enumerate}
  Similarly, we define~$\bar \genplainexec$ by removing all crash events and all
  events~$e$ of labeled with~$m(\vec v, \bot)$ such that $\finv{f}(v) \cap
  \makeset{\Ptag} = \emptyset$ from~$\genexec'$.

  \begin{claim}
    $\bar\genplainexec$ is a plain execution of~$\Lin(S)$, $\bar\genexec'$ is an
    execution of \pxes and $\toplain{\bar\genexec} \in \bar \genplainexec \bind
    I$, with $\bar f \coloneqq f \cap \bar E : \toplain{\bar \genexec'}
    \surjection \bar \genplainexec$.
  \end{claim}
  \begin{proof}[Proof of claim]
    The first two parts are obvious. The last part follows from the fact that,
    by construction, all~$\finv{\bar f}(m(\vec v, \bot))$ are $\kpo$-prefixes of
    the~$\finv{f}(m(\vec v, \bot))$, and that~$I$ is prefix-closed.
    For~$\finv{f}(m(\vec v, v))$ it follows from the fact that, if~$\finishOp$
    has finished executing, then all writes of the method have persisted.
  \end{proof}

  Write $\overline{\kcom}'$ the restriction of~$\kcom'$ to~$\bar\genexec'$.
  \begin{claim}
    The relation $\overline\kcom$ defined above is surjective on reads,
    and $\bar \genexec'$ is consistent.
  \end{claim}
  \begin{proof}[Proof of claim]
    If a read event $e$ is in $\bar \genexec'$, then in $\genexec'$:
    \[
      e':\flitWrite{p}(x,v) \xrightarrow{\;\kcom\;} e:\flitRead{p}(x,v)
      \xrightarrow{\;\kpo\;} e''
    \],
    with~$e'' \in \makeset{\Ptag}$, and because of the specification of \Flit
    $e'$ is a write which is $\klin$-before the read~$e$, therefore~$e$ depends
    on~$e''$ by transitivity and~$e' \in \makeset{\Ptag}$, and thus $e' \in \bar
    \genexec'$.

    Because \Flit defines SC executions, $\bar \genexec'$ is also consistent for
    \pxes-consistency (namely TSO).
  \end{proof}

  We can use the fact that~$I$ is assumed to implement the library~$\Lin(S)$ to
  obtain a correct execution~$\bar\genexec \refines \bar \genplainexec$.
  That is, there is a total order $\klin$ which extends~$\bar \genexec'{.}\klhb$
  and induces a sequence $r \in S$.
  Define~$e \in \genexec \iff e \in r$; and otherwise use the same $\kcom$,
  $\ksw$ and so forth from $\bar\genexec$.
\end{proof}

\section{Correctness of \Mirror}
\label{app:mirror-proof}

We consider the \Mirror library \citep{Mirror}, whose interface is that of
persistent registers. Besides the constructor, $\mirNew()$, which allocates a
new register, it provides two kinds of durable invocations: $\mirWrite(\ell,
  v)$, which writes the value $v$ at location $\ell$, and $\mirCAS(\ell, v_1,
  v_2)$, which atomically replaces the value at location~$\ell$ by~$v_2$ if it is
equal to~$v_1$ and does nothing otherwise; and one other non-durable
invocation: $\mirRead(\ell)$, which returns the value of the register at
location $\ell$.

As with \Flit, the \Mirror implementation avoids issuing a flush on reads,
albeit using a different approach: it keeps two copies of the register
contents: one in NVM and one in volatile memory. Reads access only the volatile
copy, whereas writes and CASes first write the value to NVM, flush it, and then
also write it to the volatile copy.
To ensure lock-freedom, the implementation uses sequence numbers and a
double-word (128 bits) compare-and-swap (DWCAS) to atomically update a pair of
a value and sequence number (see \cref{listing:mirror}), which is available on
\intelname.

\begin{figure}[t]
  \begin{multicols}{2}
    \begin{lstlisting}[numbers=left, language=Imp,xleftmargin=3em]
  method CAS(v_addr, expected, newval) {
    p_addr = COMPUTE_P_ADDR(v_addr);
    while (true) {
      p_seq = p_addr->seq;
      p_val = p_addr->val; *@\label{line:first-val-read}@*
      p_seq_again = p_addr->seq;

      v_seq = v_addr->seq;
      v_val = v_addr->val;
      v_seq_again = v_addr->seq;

      if (p_seq != p_seq_again ||
          v_seq != v_seq_again)
        continue;
      if (p_seq == v_seq + 1) {
        FLUSH(p_addr); FENCE();
        DWCAS(v_addr, {v_val, v_seq},
                      {p_val, p_seq});
        continue;
      }
      if (p_seq != v_seq) continue;

      if (p_val != expected) {
        expected = p_val; return false;
      }
      before = {p_val, p_seq};
      after = {vewval, p_seq+1};
      res = DWCAS(p_addr, before, after); *@\label{line:main-CAS}@*
      FLUSH(p_addr); FENCE(); *@\label{line:main-flush-fence}@*
      if (res) {
        DWCAS(v_addr, before, after); *@\label{line:success-CAS}@*
      } else {
        if (before.val == expected) continue;
        DWCAS(v_addr, {v_val, v_seq}, before); *@\label{libe:fail-CAS}@*
      }
      return res;
    }
  }

  method wr(v_addr, val) {
    p_addr = COMPUTE_P_ADDR(v_addr);
    while (!CAS(v_addr, *p_addr, val)) {}
  }

  method rd(v_addr) { return *v_addr; }
  \end{lstlisting}
  \end{multicols}\vspace{-10pt}
  \caption{The \Mirror implementation in \pxes}
  \label{listing:mirror}
\end{figure}

\subsection{Specification of the library}

\Mirror provides sequentially consistent registers with the guarantee that completed writes persist, and that
the persistence order agrees with the sequential order of the operations.

\begin{definition}
  An "execution" $\genexec$ of the library \Mirror is "correct" if there exists
  an order~$\knvo$ and a total order $\klin$ on $\genexec.E$ agreeing
  with~$\genexec.\kpo$ and~$\genexec.\klhb$ ($\genexec.\kpo \cup \genexec.\klhb
  \suq \klin$) such that:
  \begin{enumerate}
    \item $\ksw = \bigcup_{\ell \in \Loc} ([W_\ell] ; \klin ; [R_\ell])
            \setminus (\klin ; [W_\ell] ; \klin)$,
          where $W_\ell$ is the set of all writes and CASes at location $\ell$,
          and $R_\ell$ is the set of all reads and CASes at location $\ell$;
    \item if $(w, r) \in \ksw$, then $\loc(w) = \loc(r)$ and the value read by $r$ is
          the value written by $w$;
    \item $[W]; (\kpo \cup \ksw)^+; [W] \subseteq \knvo$, where $W$ is the set of all writes in $\genexec$;
    \item $\dom(\knvo; \diagonal{\Ptag}) \subseteq \makeset{\Ptag}$; and
    \item $\makeset{\Ptag} = \{ w \in W \mid \text{$w$ is a "complete operation" } \}$.
  \end{enumerate}
\end{definition}

\begin{theorem}
  The implementation in \cref{listing:mirror} is a correct
  implementation of the \Mirror library.
\end{theorem}
\begin{proof}
  Suppose given a \pxes execution~$\genexec$ and a plain
  execution~$\genplainexec'$ of \Mirror such that~$\toplain{\genexec}
  \surjection \genplainexec'$ and such that $\genexec$ is
  consistent.

  \paragraph{Operations are persisted} Define the $\Ptag$-tagged events to be
  the events~$e$ in~$\genplainexec'$ labeled with a $\mirCAS$ operation
  such that, in~$\finv{f}(e)$, the DWCAS at line~\ref{line:main-CAS} has
  succeeded and persisted.
  Clearly, because of the flush and the fence at line~\ref{line:main-flush-fence},
  all operations of the form~$\mirCAS(\ell, v_1, v_2, \text{\lstinline{true}})$
  are tagged with~$\Ptag$.

  \paragraph{Defining the~$\ksw$ relation}
  Given an event~$e$ labeled with an operation~$\mirRead(\ell, v)$
  in~$\genplainexec'$, we define the originating write of~$e$, $u(e) \in
  \genplainexec'$, as follows: consider~$r$, the corresponding read
  in~$\genexec$, and~$w$ the unique event such that~$(w, r) \in
  \genexec{.} \krf$. If~$w$ corresponds to the CAS operations of
  line~\ref{line:success-CAS}, then define $u(e) \eqdef f(w)$.
  Otherwise, let~$r'$ be the read of~\lstinline{p_addr} which precedes it
  (either line~5 or line~\ref{line:main-CAS} when the CAS is unsuccessful) and
  let~$w'$ be the write such that~$(w', r') \in \genexec{.}\krf$ and
  define~$u(e) \eqdef f(w')$.

  Now, consider a CAS event~$e \in \genplainexec'$. If it returns false, define~$u(e)
  \eqdef f(\finv{\krf}(r))$, where $r$ is the read of~\lstinline{p_addr->val}
  in line~\ref{line:first-val-read}. Otherwise, the DWCAS of
  line~\ref{line:main-CAS} succeeds,
  and we define $u(e) \eqdef f(\finv{\krf}(u))$, where $u$~is the event
  corresponding to that DWCAS.

  Finally, we define~$\ksw' \eqdef \finv{u}$, which is, by construction, the
  inverse of a function and is surjective on reads and CASes. Because the value
  written in \lstinline{v_addr->val} is the same as the value read from the
  source of the $\kcom$ edge, the value which is read matches the one which is
  written.

  \paragraph{Defining the $\klin$ order}

  We define $\klin$ by projecting along~$f$ a linearization of $\ktso$ of the
  linearization points of the operation implementations: for $\mirCAS$, it is
  the DWCAS of lines~\ref{line:main-CAS} if control-flow reaches them, and
  otherwise the read of~\lstinline{p_addr->val} in
  line~\ref{line:first-val-read}. Clearly it agrees with~$\kpo'$.

  It remains to prove that the relation~$\ksw'$ defined above reads the most
  recent write, and that it agrees with the~$\knvo' = \univproj{f}{\knvo}$
  order, where~$\univproj{f}(r) = \{ (x,y) \mid \forall u \in \finv{f}(x),\;
  \forall v \in \finv{f}(y), \; (u, v) \in r \}$.
  The first conjunct follows from the fact that all values are written using a \pxes
  DWCAS, and that all writes that are in~$P_i$ have their
  corresponding writes to~\lstinline{p_addr} persisted.
  The second conjunct follows from the fact that the linearization points of all persistent operations
   are writes to the same location, and thus their~$\ktso$
  order agrees with~$\knvo$.

  \paragraph{The relation~$[W]; (\kpo' \cup \ksw')^+; [W]$ is included in~$\knvo'$}

  We first note that given $(r, w) \in \kpo' \cap (R \times W)$, the
  reads in~$\finv{f}(r)$ are $\ktso$-before the writes in~$\finv{f}(w)$.
  Also, given $(w, r) \in \ksw' \setminus \kpo$, the read~$r'$ of
  \lstinline{v_addr} in $\finv{f}(r)$ is $\ktso$-before $\finv{\krf}(r')$ since
  they belong to different threads.
  In all paths, $r$ is $\kpo$-after a flush and a fence which is $\ktso$-after
  the write to~\lstinline{p_addr}.

  Now, let $(w, w') \in [W]; (\kpo' \cup \ksw')^*; [W]$, there exists a sequence
  of events which are related by the two relations we just discussed. Therefore
  the write to \lstinline{p_addr} in $\finv{f}(w')$ is $\ktso$-after a fence, which is
  $\ktso$-after a flush of~\lstinline{p_addr} which is $\ktso$-after the
  persistent write in~$\finv{f}(w)$.
  As such, $(w, w') \in \knvo' = \univproj{f}{\knvo}$.
\end{proof}

\subsection{Using \Mirror to Enforce Durable Linearizability}

As with \Flit, \Mirror can be used to transform a "linearizable" data-structure
into a "durably linearizable" one. Given an implementation $I$ over $\pxes$
using reads, writes and CASes, let $m(I)$ be the implementation over \Mirror
which replaces the \pxes calls with their corresponding \Mirror calls.

\begin{theorem}
  If $\libimpl{\pxes}{I}{\Lin(S)}$, then $\libimpl{\Mirror}{m(I)}{\DurLin(S)}$.
\end{theorem}

The proof is similar to the corresponding theorem of \Flit, noting that \Mirror
provides a stronger specification than \Flit.

\section{Proof of Transaction library}
\label{sec:transaction-proof}

\subsection{Proof of $\Ltrans$}

We use a module~$\makeMM{Q}$ which provides a durably linearizable queue.
One simple solution to implement this module is to take any linearizable queue
(\eg the one proved in \cite{yacovet}) and use the results of
\cref{sec:flit_and_mirror} to obtain a durably linearizable queue.
We prove the implementation~$\PTimpl$, presented in
Figure~\ref{fig:Ltrans-implementation} of the transaction library.
\begin{theorem}
  $\PTimpl$ implements the library specification~$\Ltrans$. Formally,
  $\genlib_{\pxes}, \genlib_{\mathsf{Queue}} \vdash \PTimpl : \Ltrans$.
\end{theorem}
\begin{proof}[Proof sketch]
  The idea is to define $\knvo$ on the events of~$\Ltrans$ to be~$\khb$ induced
  by the implementation-level execution graph. By well-formedness, we know all
  events are in a critical section, and we declare that an event has
  tag~$\perTag$ if the corresponding appending of COMMITTED has persisted.

  In the absence of a crash, condition (7) follows from the fact that, since
  beginnings and ends of critical sections are externally synchronized, all
  reads and writes of PT registers are related by~$\khb$. In case of a crash,
  if a read~$r$ reads from a write~$w$ with a crash in-between, we know by
  definition that~$w \in \makeset{\perTag}$, and, according the well-formedness
  condition of $\Ltrans$, that a call to recovery is~$\khb$-between the crash and~$r$.
  Since~$w \in \makeset\perTag$, the COMMITTED message has been written to the
  log and thus the write has not been undone by the recovery. Symmetrically, we
  also know that all later writes to the register have not been persisted, and
  thus the recovery procedures has undone these writes.

  Condition (8) holds for a similar reason: a read from another section is
  $\khb$-after the end of the section that wrote into the register. Conditions
  (11) and (10) which are treated later imply that the write has persisted.

  \paragraph{Global correctness} Since this library introduces it own tags, we
  can assume that the operations tagged with its own tags are its own
  operations.

  Condition (10) by definition of~$\knvo$ above.

  Condition (11): a write is persisted if the COMMITTED message of the
  corresponding critical section has been persisted. Therefore either all writes
  or none of the writes of a section persist.

  Condition (12): If $\PTEnd$ has finished, the sfence instruction has finished
  and the COMMITTED message has persisted.
\end{proof}

\subsection{Proof of $\LStrans$}

\newcommand{\lock}{\mathtt{lock}}
\newcommand{\unlock}{\mathtt{unlock}}

We consider a (volatile) lock library with the following specification: An
execution~$\genexec$ is correct if the events are totally ordered by~$\khb$ in
such a way that the induced word is of the form~$(\lock \cdot \unlock)^* \cdot
\lock^?$.

\begin{theorem}
  The implementation~$\LPTimpl$ of~$\LStrans$ is correct:
  $\genlib_{\mathsf{Lock}}, \Ltrans \vdash \LPTimpl: \LStrans$.
\end{theorem}
\begin{proof}[Proof sketch]
  The salient part of the proof is establishing that the library TC is used
  according to its well-formedness specification. According to
  Definition~\ref{def:well-formed-execution}, we consider an implementation
  execution $\genexec$ which has an immediate prefix~$\bar\genexec \prefimm
  \genexec$ which is correct.
  The important case is when the event added in $\genexec$ compared
  to~$\bar\genexec$ is a call to~$\PTBegin$ or~$\PTEnd$ which is part of the
  implementation of a call to $\LPTBegin$ or~$\LPTEnd$ respectively.
  In that case, the correctness of~$\bar\genexec$ restricted to the Lock library
  implies that two LPT critical sections are related by $\khb$. Therefore, the
  calls to the PT library are externally synchronized.

  The rest of the proof consists in using unchanged the corresponding properties
  of the PT library: We consider a correct execution~$\genexec \in
  \Exec{\genlib_{\makeMM{Lock}}, \Ltrans, \libtags{\Ltrans}}$ and a corresponding
  locally correct execution~$\genexec \in \Exec{\LStrans, \Ltrans,
  \libtags{\Ltrans}}$ and we need to prove that~$\genexec$ is correct with
  respect to the global specification of~$\Ltrans$. This is easy to see that this
  follows from the correctness of~$\genexec$ with respect to the same global
  specification.
\end{proof}

\subsection{Counter}

\begin{lstlisting}[language=Imp]
  method NewCounter() = PTAlloc()

  method CounterInc(c) = let v = Read(c) in PTWrite(v+1)

  method CounterRead(c) = PTRead(c)
\end{lstlisting}

An execution is~$\Counterlib$-correct if each call to~$\mathsf{CounterRead(c)}$
returns the number of~$\mathsf{CounterInc(c)}$ which are~$\khb$-before it.
The proof of correctness is simple, for example by using condition (7) of the
specification of~$\Ltrans$.

Note that in the proof, the implementation graph~$\genexec$ we consider is
\emph{not} well-formed, since there are no begin/end calls!
However, we know that for any well-formed context in which~$\Counterlib$ is
used, the library~$\Ltrans$ will also be used according to well-formedness, and
as such we can use the fact that~$\genexec$ is correct.

\end{document}